\documentclass{aa}

\usepackage{graphicx}
\usepackage{txfonts}
\usepackage{amsmath}
\usepackage{amssymb}
\usepackage{wasysym}
\usepackage{color}
\usepackage{epsfig}
\usepackage{longtable}
\usepackage{pdflscape}
\usepackage{mathtools}
\usepackage{float}
\usepackage{footnote}
\usepackage{epstopdf}
\usepackage{units}
\usepackage{enumerate}
\usepackage{pdflscape}

\DeclareRobustCommand{\ion}[2]{\textup{#1\,\textsc{\lowercase{#2}}}}
\DeclareRobustCommand{\teff}{T_{\mathrm{eff}}}
\DeclareRobustCommand{\logg}{\log g}
\DeclareRobustCommand{\mh}{\mathrm{[M/H]}}
\DeclareRobustCommand{\vmic}{\varv_\mathrm{mic}}

\DeclareRobustCommand{\ispec}{\texttt{iSpec} }
\DeclareRobustCommand{\kms}{$\mathrm{km s}^{-1}$}
\DeclareRobustCommand{\vr}{$v_\mathrm{r}$}

\usepackage[acronyms,style=super,nonumberlist,nogroupskip]{glossaries}

\newacronym[plural={OCs}, \glsshortpluralkey={OCs}]{OC}{OC}{Open Cluster}
\newacronym[]{SNR}{S/N}{signal-to-noise ratio}
\newacronym[]{GBS}{GBS}{\emph{Gaia} FGK Benchmark Stars}
\usepackage[]{hyperref}

\begin{document} 

  \title{Differential abundances of open clusters and their tidal tails: chemical tagging and chemical homogeneity\footnote{Thanks to observations at Telescope Bernard Lyot and data retrieved from the archives: ESO, TNG, FIES, ESPaDOnS and NARVAL}}
  \titlerunning{Differential abundances of nearby Open Clusters and their tidal tails}
  \author{L. Casamiquela\inst{1} \and Y. Tarricq\inst{1} \and C. Soubiran\inst{1} \and S. Blanco-Cuaresma\inst{2} \and P. Jofr\'e\inst{3} \and U. Heiter\inst{4} \and M. Tucci Maia\inst{3}}

  \institute{Laboratoire d’Astrophysique de Bordeaux, Univ. Bordeaux, CNRS, B18N, allée Geoffroy Saint-Hilaire, 33615 Pessac, France\\
  \email{laia.casamiquela-floriach@u-bordeaux.fr}
  \and
  Harvard-Smithsonian Center for Astrophysics, Cambridge, MA 02138, USA
  \and
  N\'ucleo de Astronom\'ia, Facultad de Ingenier\'ia y Ciencias, Universidad Diego Portales, Av. Ej\'ercito 441, Santiago, Chile
  \and
  Observational Astrophysics, Department of Physics and Astronomy, Uppsala University, Box 516, 75120 Uppsala, Sweden}

  \date{Received ; accepted}

   \abstract
   {Well studied Open Clusters (OCs) of the Solar neighbourhood are frequently used as reference objects to test galactic and stellar theories. For that purpose their chemical composition needs to be known with a high level of confidence. It is also important to clarify if each OC is chemically homogeneous and if it has a unique chemical signature.}
   {The aims of this work are (1) to determine accurate and precise abundances of 22 chemical species (from Na to Eu) in the Hyades, Praesepe and Rupecht~147 using a large number of stars at different evolutionary states, (2) to evaluate the level of chemical homogeneity of these OCs, (3) to compare their chemical signatures.}
   {We gathered $\sim$800 high resolution and high signal-to-noise spectra of $\sim$100 members in the three clusters, obtained with the latest memberships based on Gaia DR2 data. We build a pipeline which computes atmospheric parameters and strictly line-by-line differential abundances among twin stars in our sample. With this method we are able to reach a very high precision in the abundances (0.01-0.02 dex in most of the elements).}
   {We find large differences in the absolute abundances in some elements, which can be attributed to diffusion, NLTE effects or systematics in the analysis. For the three OCs, we find strong correlations in the differential abundances between different pairs of elements. According to our experiment with synthetic data, this can be explained by some level of chemical inhomogeneity.
   We compare differential abundances of several stars from the Hyades and Praesepe tails: the stars that differ more in chemical abundances also have distinct kinematics, even though they have been identified as members of the tail.}
   {It is possible to obtain high precision abundances using a differential analysis even when mixing spectra from different instruments. With this technique we find that the Hyades and Preasepe have the same chemical signature when G dwarfs and K giants are considered. Despite a certain level of inhomogeneity in each cluster, it is still possible to clearly distinguish the chemical signature of the older cluster Ruprecht~147 when compared to the Hyades and Praesepe.}

   \keywords{ (Galaxy:) open clusters and associations: individual: Hyades, NGC 2632, Ruprecht 147--
   Stars: abundances--
   Techniques: spectroscopic}

   \maketitle

\vspace{0.5cm}

\section{Introduction}

\glspl{OC} of different ages and chemical compositions are ideal to test on star formation and evolution theories and have long been used to better understand the history of the Galactic disc. Several spectroscopic surveys dedicate a significant observing time to \glspl{OC} such as \emph{Gaia}-ESO survey \citep{Gilmore+2012,Randich+2013}, APOGEE \citep{Majewski+2017}, OCCASO \citep{Casamiquela+2019}, among others. In these surveys, \glspl{OC} provide fundamental material for calibrating the stellar parameters, in particular, the dependencies of abundances as a function of stellar parameters \citep{Jofre+2019}. 

\glspl{OC} have long been thought to be chemically homogeneous \citep[e.g.,][]{Friel+2002}, from the hypothesis that the cloud from which the cluster was formed was uniformly mixed. Observed abundance dispersions are typically around 0.05 dex, usually at the same level of the measurement uncertainties. By using a strictly line-by-line differential analysis method, such uncertainties can be lowered to better assess the homogeneity of \glspl{OC}.
Differential chemical abundance analysis has been mainly used to analyze abundance variations among solar twins \citep[e.g.][]{Melendez+2009b,Nissen2015,TucciMaia+2016, Mahdi+2016}. Applications in other contexts can also be found, such as studies of nearby stars, globular or open clusters, and benchmark stars \citep[e.g.,][]{Heiter+2003,Yong+2013,Onehag+2014,Jofre+2015,Hawkins+2016b,Liu+2019}.
By using a reference star with stellar parameters close to the program stars, this method allows one to reach a very high precision in abundances (on the order of 0.01 dex) because it minimizes the uncertainty coming from errors in the characterization of the spectral lines.
In particular, \citet{Liu+2016} and \cite{Spina+2018} analyzed 16 and 5 solar analogs, members of the Hyades and the Pleiades, respectively, to show chemical inhomogeneities at the level of 0.02 dex.

In this paper, we investigate how the line-by-line differential method can be applied to stars over a larger range of evolutionary states. This is important because distant G dwarfs are usually not observable at high spectral resolution because they are too faint for obtaining high \gls{SNR} spectra with current spectrographs. We thus propose intrinsically brighter stars, such as clump giants and F dwarfs in nearby \glspl{OC} to be used as reference objects for abundance studies involving distant targets. By measuring differential abundances with respect to stars of similar evolutionary state in local \glspl{OC} we expect to see subtle variations of chemical composition in the OC population. This methodology is also well suited to further develop the concept of chemical tagging that aims to identify stars of common origin through their abundances. With our high precision differential abundances, we also measure the level of chemical homogeneity in \glspl{OC}.

As the first step to establish a list benchmark \glspl{OC} with well-characterized chemical signature, we focus on three nearby objects: the Hyades, NGC~2632 (Praesepe) and Ruprecht 147. The Hyades and NGC~2632 have been considerably observed in the past \citep[e.g.][among many others]{Gebran+2010,Boesgaard+2013,Gossage+2018}. Both appear to have similar in age, between 600 and 800 Myr, and metallicity $\sim+0.15$ dex. Many high-quality spectra are available in public archives for further analyses. Ruprecht~147 has been surprisingly less observed although it is highly interesting, being the nearest OC older than 1 Gyr. One recent study provides an analysis of its chemical composition \citep{Bragaglia+2018} after \emph{Gaia} DR2, retrieving a solar metallicity. The three clusters have clump giants, excellent targets for spectroscopy because of their brightness and sharp lines allowing precise radial velocity and abundance determinations. Their population of FGK dwarfs is also easily observable at high resolution with 2--4 m class telescopes. In this study we take advantage of new assessments of membership probabilities for stars in the fields of these clusters, dramatically improved thanks to \emph{Gaia} DR2 \citep{GaiaCollaboration+2018}. With spectra of our own observations and from public archives we provide high precision absolute and differential abundances for an unprecedented number of stars in each cluster, up to large distances from the cluster's center, including their extended halo and tidal tails.

The paper is organized as follows: the selection of the target stars and the observational material is in Sect.\ref{sec:observmat}, the method used and the explanation of the pipeline used to perform all computations is detailed in Sect.\ref{sec:method}, the membership refinement using total Galactic velocities is explained in Sect.\ref{sec:galvel}. Sect.\ref{sec:results} includes the results of the spectroscopic analysis: atmospheric parameters and chemical abundances of the cluster stars and the analyzed \gls{GBS}. In Sect.\ref{sec:diff_abund} we detail the computation of the differential chemical abundances and we study their precision, the chemical signature of the stars in the tidal tails of the clusters, the homogeneity of the three clusters and the possibility of chemical tagging.

\section{Observational material}\label{sec:observmat}

\subsection{Cluster and star selection from \emph{Gaia} DR2}

As starting point we used the memberships lists provided for known OCs by \citet{GaiaCollaborationB+2018} and by \citet{Cantat-gaudin+2018}, who made use of \emph{Gaia} DR2 astrometry. We focus on three nearby evolved OCs for which we gathered many high-resolution spectra of stars at different evolutionary stages, either from our own observations or from public archives: the Hyades (Melotte~25), Praesepe (NGC~2632), and Ruprecht~147.
Their distances, ages, and the number of stars for which high-resolution spectra are available are listed in Table~\ref{tab:clusters}.

\begin{table}
  \caption{\label{tab:clusters}Properties of the \glspl{OC} studied in this work. We indicate the cluster distance $D$ \citep[from \emph{Gaia} DR2 parallaxes,][]{GaiaCollaborationB+2018}, and ages from \citet[][D02]{Dias+2002}, \citet[][K13]{Karchenko+2005},  \citet[][GC18]{GaiaCollaborationB+2018}, and \citet[][B19]{Bossini+2019}. The number of stars with high resolution spectra is given in the rightmost column.}
\begin{centering}
\setlength\tabcolsep{2pt}
\begin{tabular}{lrrcccccc}
\hline
Cluster     & $D$ [pc] & \multicolumn{4}{c}{$\log$\,(age\,[yr])} &Num.   \\
            &          & D02  & K13  & GC18  & B19               &stars  \\
\hline
Hyades      &  48      & 8.90 &   -  & 8.90  & -                 &  62   \\
NGC~2632    & 186      & 8.86 & 8.92 & 8.85  & 8.87              &  22   \\
Rup~147     & 284      & 9.40 & 9.33 & 9.30  & -                 &  24   \\
\hline
\end{tabular}
\end{centering}
\end{table}

The stars to be spectroscopic targets were selected according to the available membership information and their positions in the color-magnitude diagram.
\begin{itemize}
\item We used the list of members from \citet{Cantat-gaudin+2018,GaiaCollaborationB+2018} without any restriction on the membership probability. We included as part of our targets the stars found as members of the Hyades tails \citep{Roser+2019,Meingast+2019}, and the NGC~2632 tails \citep{Roser+2019b}. In the case of NGC~2632 and Ruprecht~147 we added stars (up to a distance $d\sim150$ pc from the centre) which have $(U,V,W)$ values\footnote{Computed using \emph{Gaia} DR2 radial velocities} compatible with the cluster (up to $2.5$ \kms\ w.r.t. the mean cluster velocity) following the methodology of \citet{Meingast+2019}.
\item In the case of NGC~2632 and Ruprecht~147, we also used \emph{Gaia} DR2 information on radial velocity. We discarded those stars which had a radial velocity different by more than $1.2$MAD\footnote{Median Absolute Deviation} from the median value of all stars. The stars without radial velocity information were kept.
\item For a precise spectroscopic characterization with our method we require giant stars or dwarfs with temperatures within the range $6500$~K$\lesssim \teff \lesssim5000$~K. Hotter dwarfs usually have higher rotational velocities depending on age \citep{Nielsen+2013}, and cooler stars have more crowded spectra due to molecular bands, giving less precise abundances. We used the PARSEC \citep{Bressan+2012} isochrones to determine the color range in \emph{Gaia} bands that correspond to these temperature limits ($0.6\lesssim Bp-Rp\lesssim 1.1$).
\item Dwarfs that were clearly located out of the main sequence in the color-absolute magnitude diagram were excluded to avoid binaries or possible contamination by non-members.
\end{itemize}

\subsection{Spectra}

Using the selection criteria described in the previous subsection we selected more than 467 candidate stars to be studied spectroscopically. These close clusters have been studied by previous authors, and so we expected to find spectroscopic data for a large fraction of the selected stars. We queried the available public archives searching for high resolution spectra ($R=\lambda/\Delta\lambda\gtrsim45\,000$). We did not put any restriction on \gls{SNR} \emph{a priori}, so in many cases, we retrieved low \gls{SNR} spectra of the same star and instrument that can be co-added to reach a \gls{SNR} of $\sim$50.

We retrieved, reduced, and calibrated spectra from the following instruments\footnote{We also retrieved spectra from the SOPHIE archive. However, we encountered several problems in recovering atmospheric parameters from SOPHIE spectra, and hence we did not include them in this work.}:
\begin{itemize}
\item UVES: a cross-dispersed echelle spectrograph installed at the second Very Large Telescope (VLT) unit, at Paranal Observatory. It covers part of the optical spectral region with a resolution $R\sim45\,000$ or above, depending on the setup. We selected the setups according to the wavelength range of our line list, those centered on 580 and 564~nm. We retrieved the spectra using the ESO Phase 3 spectral data webpage\footnote{\url{http://archive.eso.org/wdb/wdb/adp/phase3_spectral/form}}.
\item FEROS: a high resolution ($R\sim48\,000$) echelle spectrograph providing an almost complete spectral coverage from 350 to 920 nm. It is installed at the 2.2m MPG/ESO telescope at La Silla Observatory. We retrieved the spectra using also the ESO Phase 3 archive.
\item HARPS: a very high-resolution spectrograph ($R\sim115\,000$) installed at the 3.6m telescope at La Silla Observatory. The instrument was designed to obtain very high accuracy in radial velocity. The spectral range covered is 380 to 690~nm. We retrieved the spectra using also the ESO Phase 3 archive.
\item HARPS-N: an instrument with very similar capabilities to HARPS ($R\sim115\,000$, 380 to 690 nm) attached to the Telescopio Nazionale Galileo (TNG) at the Observatory El Roque de Los Muchachos. We used the TNG archive\footnote{The archive (\url{http://archives.ia2.inaf.it/tng/faces/search.xhtml?dswid=-9619}) does not allow an automatic search for a large number of stars, so we queried the stars of Rup 147 that we knew in advance had been observed by \citet{Bragaglia+2018}.} to retrieve the data.
\item FIES: a cross-dispersed high-resolution echelle spectrograph with a spectral resolution of $R\sim67\,000$ and a spectral range coverage from 370 to 910 nm. It is attached to the Nordic Optical Telescope (NOT) at the Observatory El Roque de Los Muchachos. We used an exhaustive list of the public FIES spectra provided by the staff to cross-match with our target stars and retrieve the spectra.
\item ESPaDOnS: a high resolution ($R\sim68\,000$ -- $81\,000$ depending on the configuration) spectropolarimeter covering a spectral range of 370 to 1050~nm, mounted at the Canada-French-Hawaii Telescope (CFHT) at the Mauna Kea Observatory. We cross-matched our targets with a list of ESPaDOns public observations provided by the staff to retrieve the reduced spectra.
\item NARVAL: a twin of ESPaDOnS installed at the Telescope Bernard Lyot (TBL) atop of the Pic du Midi observatory. We used the same strategy as for ESPaDOnS to retrieve the spectra.
\item ELODIE: it was an echelle spectrograph ($R\sim42\,000$, 390 to 680 nm) installed at the Observatoire de Haute-Provence (OHP) 1.93m telescope until 2006. We used its dedicated archive\footnote{\url{http://atlas.obs-hp.fr/elodie/}}.
\end{itemize}

Additionally, we performed our own observing programs with NARVAL during two semesters (2018B and 2019A), where we observed a total of 25 stars.

In total, we collected 848 spectra corresponding to 108 different stars: 62, 22 and 24 from the Hyades, NGC~2632, and Ruprecht~147, respectively. In the cases where different spectra from the same instrument corresponded to the same star, we co-added them to reach a higher \gls{SNR}. Several stars were observed also with different instruments, in this case, the spectra were treated independently for comparison purposes.

\begin{figure}
\centering
\includegraphics[width=0.5\textwidth]{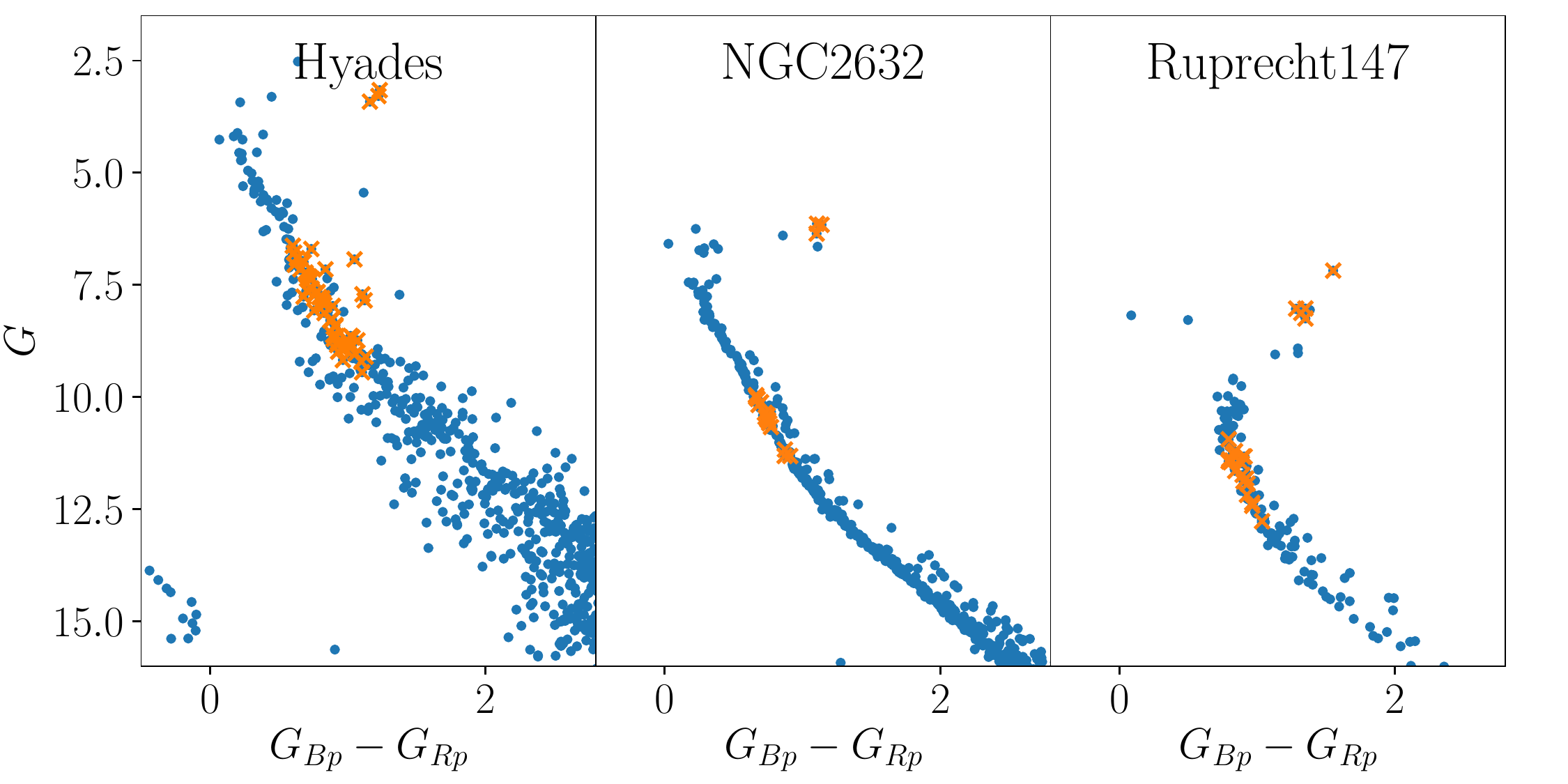}
\caption{Color-magnitude diagrams of the stars of the three clusters. In blue we plot the initial list of members from the studies indicated in the text. Target stars with retrieved spectra are plot with orange crosses.}\label{fig:HRdiag}
\end{figure}

We show in Fig.~\ref{fig:HRdiag} a color-magnitude diagram of the cluster members, indicating the targets with spectra.

\section{Method}\label{sec:method}
We used the public spectroscopic software \ispec \citep{BlancoCuaresma+2014,BlancoCuaresma2019} to analyze the spectra. This is a Python code designed to perform operations on stellar spectra and to compute radial velocities, atmospheric parameters, and individual chemical abundances using different available atmospheric models and radiative transfer codes.

We employed the synthetic spectral synthesis method to compute atmospheric parameters and chemical abundances using the radiative transfer code SPECTRUM \citep{Gray+1994}, the MARCS\footnote{http://marcs.astro.uu.se/} atmospheric models \citep{Gustafsson2008}, and the solar abundances by \citet{Grevesse+2007}. We used the line list from the \emph{Gaia}-ESO survey \citep{Heiter+2015b,Heiter+2019}. The spectral fitting was done comparing the observed fluxes weighted by their uncertainties with a synthetic spectrum for a set of spectral features. Atmospheric parameters and chemical abundances were varied in two separate steps until convergence was reached using a least-squares algorithm.

\subsection{Pipeline}\label{sec:pipeline}
We adapted the pipeline used in \citet{BlancoCuaresma+2018}, using a general workflow described below.

In a first pre-processing step each spectrum is cut to a restricted common wavelength range (480 to 680 nm) and downgraded to a common resolution ($45\,000$) to be analyzed homogeneously. Heliocentric radial velocities are computed and the different spectra from the same star and instrument are co-added to reach high \gls{SNR}. The radial velocity of the co-added spectrum is determined from cross-correlation with a high \gls{SNR} solar spectrum from NARVAL. Strong telluric absorption or emission lines are identified and masked using a telluric line list. The spectrum is normalized to the continuum using quadratic splines, with nodes distributed along the spectrum at every 5 nm. The continuum level is found using a median and maximum filter, to account for the absorption lines.

The atmospheric parameters $\teff$, $\logg$, $\mh$, and $[\alpha\mathrm{/M}]$, and the microturbulence parameter $\vmic$ are inferred for each spectrum using spectral synthesis fitting, in a non-differential way. We used the master line list by \citet{BlancoCuaresma2019} and also the wings of H$\alpha$, H$\beta$, and the Mg~I b triplet lines.

As for the broadening effects, the projected equatorial rotational velocity $\varv\sin i$, the macroturbulence parameter, and the spectral resolution are degenerate and difficult to disentangle. We applied the strategy described in \citet{BlancoCuaresma2019}: we used a fixed value for $\varv\sin i$ of 1.6 \kms, the macroturbulence was computed with the empirical relation used in the \emph{Gaia}-ESO Survey (M. Bergemann and V. Hill, priv. comm.), and only the spectral resolution was let free, accounting for all broadening effects. 

Absolute chemical abundances of individual lines were measured using the atmospheric parameters fixed to the values resulting from the previous step.
To derive chemical abundances, an additional cleaning of lines was done discarding systematically discrepant lines of each element in most of the stars. This was done for three groups of stars according to evolutionary state: K giants ($\logg<3.5$), G dwarfs ($5000<\teff<5900$ K), and F dwarfs ($5900<\teff<6400$ K)\footnote{These limits in temperature do not correspond to the exact definition of spectral types. However, the two groups are dominated by the G and F type stars, respectively. We use this nomenclature throughout the paper for simplicity.}. For elements with few measured spectral lines, we used the flags included in the \emph{Gaia}-ESO line list, indicating the reliability of the atomic data and their degree of blending. We tested that the obtained lines were consistent using stars observed with different instruments. The final selection of lines is in Table~\ref{tab:linelist}.

In a final step, differential abundances were calculated line by line, subtracting the abundance values of a chosen reference star (see Sect.~\ref{sec:diff_abund}). This procedure was restricted to those lines present in the reference star.

\begin{table}
\caption{Selection of lines used to compute chemical abundances. We indicate for each line: element, wavelength ($\lambda_{\rm peak}$), atomic information ($\log gf$ and Excitation Potential -EP-), and type of stars. The full table is available at CDS. }\label{tab:linelist}
\centering
\begin{tabular}{lcccc}
 \hline
Element     & $\lambda_{\rm peak}$ &$\log gf$& EP     & Stars\\
\hline
\ion{Ca}{I} & 534.9465              & -0.310 & 2.7090 & CD,WD\\
\ion{Ca}{I} & 526.1704              & -0.579 & 2.5210 & WD,G \\
\hline
\end{tabular}
\end{table}

\section{Galactic velocities: membership refinement}\label{sec:galvel}

We used radial velocities computed by \ispec to identify kinematic outliers in each cluster. Such stars could be no (or less reliable) members, or spectroscopic binaries, which we want to remove from our sample to retrieve chemical abundances of member stars whose chemical pattern reflects the composition of the gas cloud, and not anomalies due to binary interaction.
We find compatible radial velocities among different spectra of the same star. Several stars have significantly different radial velocities w.r.t. the rest of the cluster stars. This is expected because of the projection effects on the sky in these nearby objects, in particular for the Hyades and NGC~2632.

We identified 9 stars with large uncertainties in the radial velocity ($>1.5$ \kms) which tend to have large FWHM of the spectral features ($\gtrsim30$ \kms). We list them in the upper part of Table~\ref{tab:outliers}. These correspond to warm stars ($\teff\gtrsim6300$ K), which possibly rotate more rapidly than solar-type stars \citep{Nielsen+2013}. The uncertainties in the abundances of all these stars are significantly larger than for the rest of our sample. We exclude all of them for the analysis in the next subsections because it will not be possible to retrieve reliable abundances with our employed procedure of differential analysis. In particular, we remark Gaia DR2 43789772861265792 (V471 Tau) has a FWHM 100 times larger than the rest of the stars. This is one of the giant stars in the Hyades which is a known eclipsing binary.

We used \emph{Gaia} DR2 proper motions and positions, and the derived radial velocities to compute Galactic 6D coordinates ($X,Y, Z, U, V, W$) using \texttt{pygaia}\footnote{\url{https://github.com/agabrown/PyGaia}}. For the stars of each of the three clusters, we applied a $3\sigma$ rejection until a dispersion in each velocity coordinate of $<2.5$ \kms\ was reached, corresponding to the typical dispersion expected for a cluster \citep{Riedel+2017}.
We identified three discrepant stars in the Hyades, and three in Ruprecht~147. See the lower part of Table~\ref{tab:outliers} for a summary of these stars. We removed them from the next subsections. Several of the outliers correspond to the preceding and trailing tidal tails of the Hyades, these are analyzed in Sect.~\ref{sec:tails}. One of the Hyades outliers has been previously identified in the literature as spectroscopic binary. From the outliers identified in Ruprecht~147, two do not have previous measurements in the literature. The other has two previous measures, discrepant with our value and among them. It is a possible spectroscopic binary.

\begin{table}
\caption{Stars with large \vr uncertainty (upper part), and identified outliers using total velocities (lower part). We indicate the cluster and the \emph{Gaia} DR2 source id. SB=spectroscopic binary).}\label{tab:outliers}
\centering
\setlength\tabcolsep{3pt}
\begin{tabular}{lcc}
 \hline
 Cluster & Star                &  Comments                   \\
 \hline
 Hyades  & 43789772861265792   & EB (V471 Tau)               \\
 Hyades  & 144130516816579200  & Tidal tails                 \\ 
 Hyades  & 3312837025641272320 & Tidal tails                 \\ 
 Hyades  & 149313099234711680  &                             \\ 
 Hyades  & 3305871825637254912 &                             \\ 
 Hyades  & 3314212068010812032 &                             \\ 
 Hyades  & 3393284752392701312 &                             \\ 
 NGC~2632& 1918687411545919232 & Tidal tails                 \\ 
 NGC~2632& 661419259867455488  &                             \\ 
\hline
 Hyades  & 2495442626804315392 & Tidal tails                 \\
 Hyades  & 3380479015342121600 & Tidal tails                 \\
 Hyades  & 145293181643038336  & SB$^1$                      \\
 Rup~147 & 4087807180650392832 &                             \\
 Rup~147 & 4087786874044570880 &                             \\
 Rup~147 & 4184144534049662720 & Lit. discrepant \vr$^2$, SB?\\
 \hline
\end{tabular}
\flushleft$^1$\citet{White+2007}\\
$^2$\citet{GaiaCollaboration+2018,Curtis+2013}
\end{table}

\vspace{0.2cm}
The median and MAD of the cluster radial velocities determined with our set of high resolution spectra (excluding outliers from Galactic velocities) are $39\pm2$ \kms\ for the Hyades (59 stars), $34.4\pm0.8$ \kms\ for NGC~2632 (22 stars), and $41.4\pm0.5$ \kms\ for Rup~147 (21 stars). These values are in good agreement with the mean radial velocities derived by \citet{GaiaCollaborationB+2018} and \citet{Soubiran+2018} from \emph{Gaia} DR2 data.

\section{Results of the spectroscopic analysis}\label{sec:results}

\subsection{Atmospheric parameters}\label{sec:APs}
\begin{figure}
\centering
\includegraphics[width=0.5\textwidth]{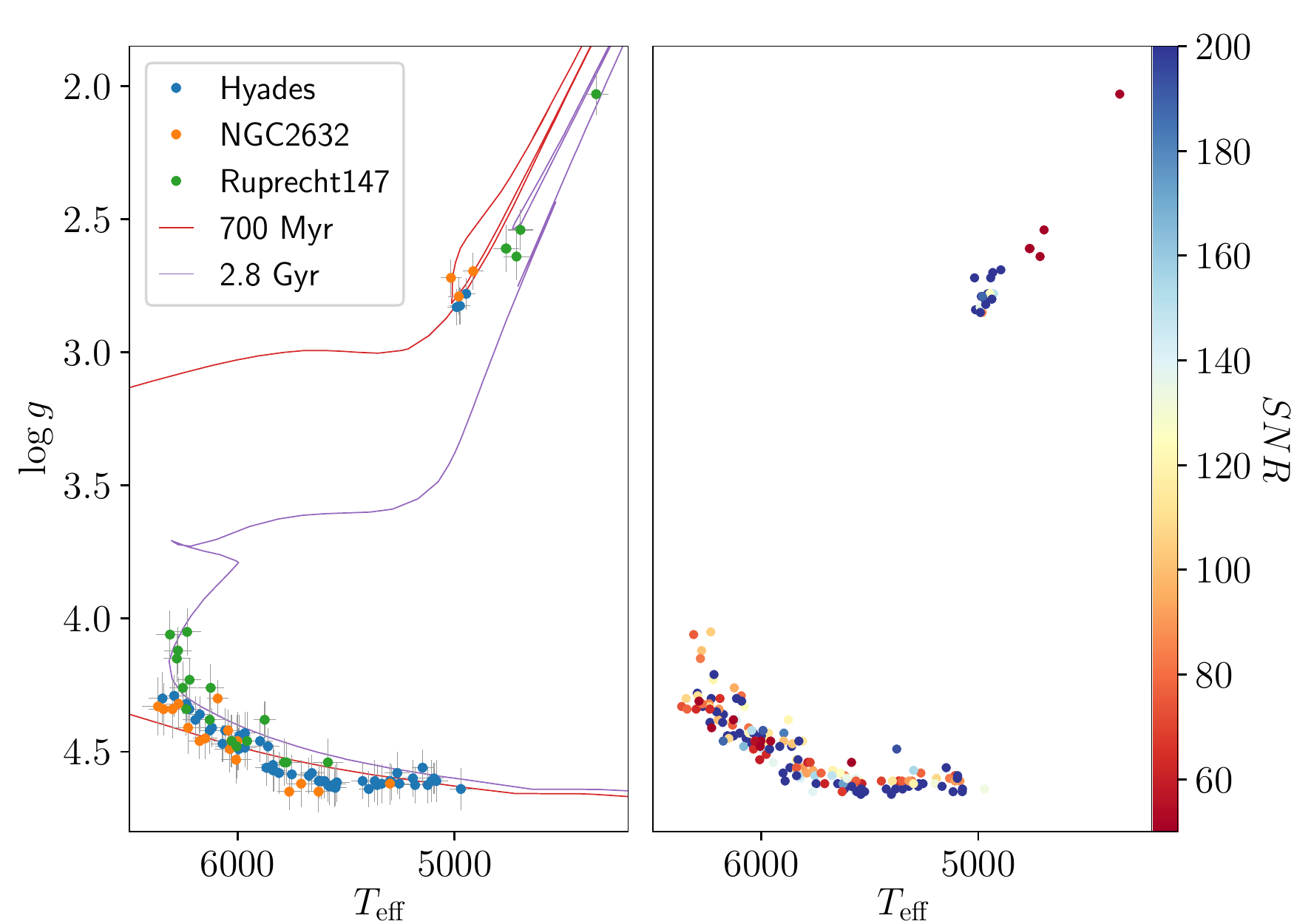}
\caption{HR diagram showing the $\teff$ and $\logg$ values resulting from the analysis of the target stars. Left: stars are colored according to the cluster, using median values for the stars with spectra from different instruments. We overplot two isochrones representative of the ages of the clusters. Right: all spectra colored by \gls{SNR}.}\label{fig:AP}
\end{figure}

We show in Fig.~\ref{fig:AP} the $\teff$ and $\logg$ resulting from the analysis of the bona fide member stars. The uncertainties are computed adding the quoted uncertainty delivered by \texttt{iSpec} to the mean dispersion obtained from the comparison of the GBS (see Table~\ref{tab:GBS_differences}). We overplot two isochrones\footnote{PARSEC isochrones \citep{Marigo+2017}, \url{http://stev.oapd.inaf.it/cgi-bin/cmd}} representative of the ages of the Hyades and Praesepe (700 Myr) and Ruprecht~147 (2.8 Gyr). One can see three clear main sequences and red clumps corresponding to the three clusters. The locus of the main-sequence turn-off and of the giants in Ruprecht~147 differ slightly from the other two, because of its older age.
One giant star in Ruprecht~147 is brighter and cooler than its red clump (Gaia DR2 4183949198935967232). This star was identified as a red giant branch star by \citet{Carlberg2014}. 

\begin{figure}
\centering
\includegraphics[width=0.5\textwidth]{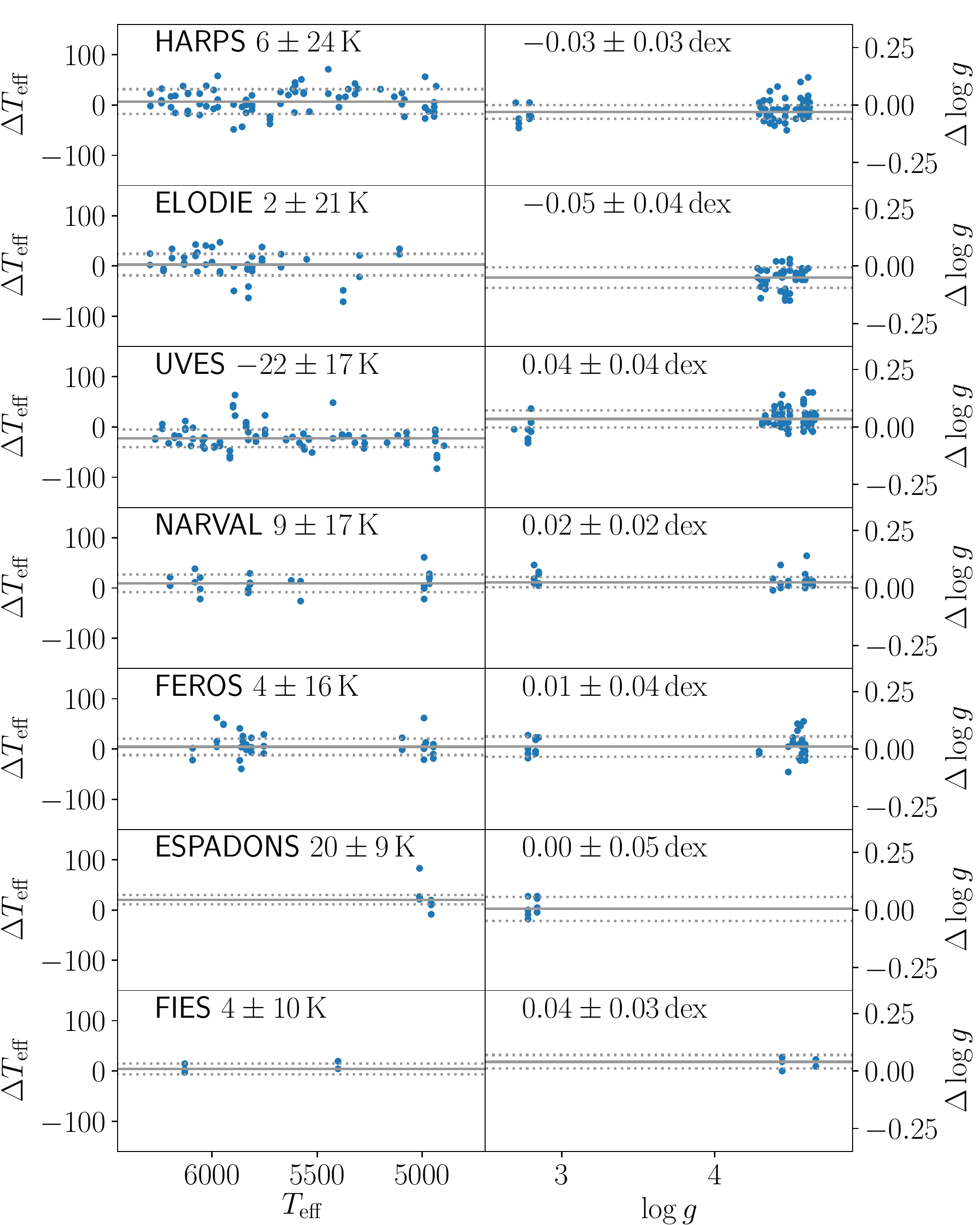}
\caption{Differences in $\teff$ (left column) and $\logg$ (right column) for stars observed with different instruments. For each instrument we show the difference as: instrument -- others, as a function of the value obtained for that instrument. We indicate the median and MAD difference in each panel.}\label{fig:AP_repeated}
\end{figure}

We used the sample of stars which were observed with more than one instrument to check the internal consistency of the atmospheric parameters. In total, 44 stars have several observations. In Fig.~\ref{fig:AP_repeated} we show, for each instrument, the difference in $\teff$ and $\logg$ between the value obtained for the spectrum of that instrument and the value for the same star observed with other instruments. The median offsets in $\teff$ and $\logg$ are lower than 22 K and 0.05 dex, respectively, and the dispersions (MAD) are of a similar level. Only for $\teff$ in ESPaDOnS we obtain an offset which is significantly larger than the dispersion, but in this case, we have a very small number of stars.

\subsection{Chemical abundances}

We compute LTE chemical abundances as explained in Sec.~\ref{sec:method} for 22 chemical species of all nucleosynthetic channels:
\ion{Na}{I}, 
\ion{Al}{I}, 
\ion{Mg}{I}, 
\ion{Si}{I}, 
\ion{Ca}{I}, 
\ion{Sc}{II}, 
\ion{Ti}{I}, 
\ion{Ti}{II}, 
\ion{V}{I}, 
\ion{Cr}{I}, 
\ion{Mn}{I}, 
\ion{Fe}{I}, 
\ion{Fe}{II}, 
\ion{Co}{I}, 
\ion{Ni}{I}, 
\ion{Cu}{I}, 
\ion{Y}{II}, 
\ion{Ba}{II}, 
\ion{La}{II}, 
\ion{Ce}{II}, 
\ion{Nd}{II},
\ion{Eu}{II}. We have selected the lines computed of each element discarding systematically discrepant lines, as explained in Sec.~\ref{sec:pipeline}. Several elements could only be analyzed in certain spectral types (La, Ce), according to the line selection made in Sect.~\ref{sec:pipeline}.

\begin{figure*}
\centering
\includegraphics[width=\textwidth]{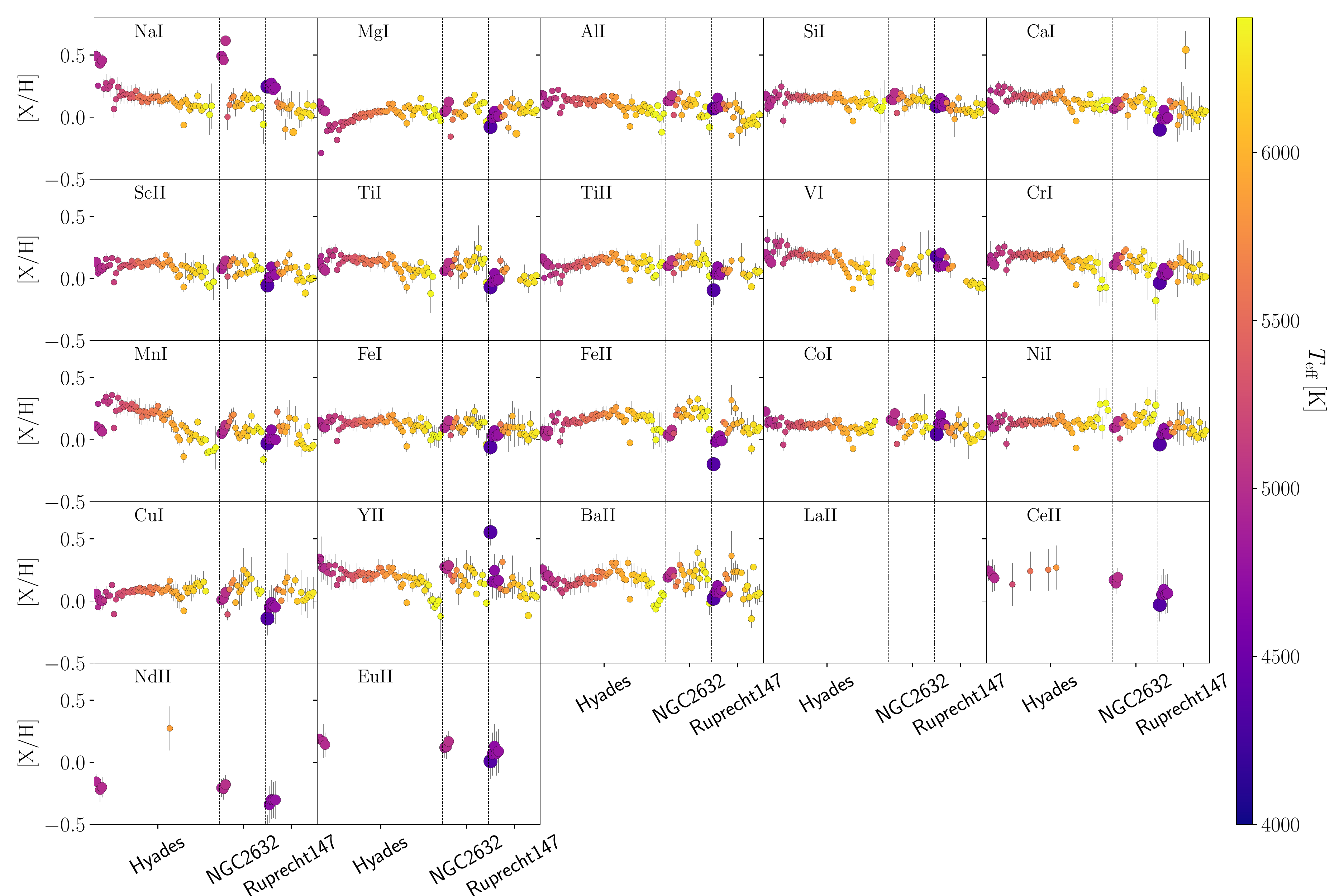}
\caption{Abundances [X/H] of each chemical species computed for the stars in the analyzed OCs. The color codes the effective temperatures and the size represents the surface gravity (larger sizes correspond to giant stars). Vertical lines separate the stars of the three clusters.}\label{fig:absolute_abus}
\end{figure*}

From the absolute\footnote{Absolute abundance is defined as $A_X = \log\left(\frac{N_X}{N_H}\right) + 12$ where $N_X$ and $N_H$ are the number of absorbers of the element atoms, and of hydrogen, respectively.}
chemical abundances retrieved from \ispec, we compute bracket abundances with respect to the Sun ([X/H]) using the median solar abundance obtained with the 9 available spectra from the Sun analyzed within the \gls{GBS} (see next subsection).
The results are plotted in Fig.~\ref{fig:absolute_abus} for the stars in each cluster sorted by $\teff$. Stars observed several times are represented by the mean value of the element abundance, and the squared sum of the standard deviation and the mean of the quoted errors was assigned as the uncertainty in abundance. Abundances with large uncertainties ($>0.2$ dex) are rejected in the discussion and the plot, this lowers considerably the number of stars in heavy elements (\ion{Ce}{II}, \ion{Nd}{II} and \ion{Eu}{II}).

Several chemical species such as \ion{Na}{I}, \ion{Mg}{I}, \ion{V}{I}, \ion{Mn}{I} and \ion{La}{II}, show a significant gradient in abundance with effective temperature and surface gravity. These differences can be due to several effects:
\begin{enumerate}
\item Non-LTE effects are expected to be large in some of these elements, such as Na \citep{Lind+2011} up to 0.5 dex, and Mn \citep{Bergemann+2019} up to 0.4 dex.
\item It can also be due to a change of chemical abundances in the stellar atmosphere depending on the evolutionary stage. Several studies have found significant abundance variations among stars across different evolutionary phases in old OCs \citep[such as M67, see e.g.][]{Souto+2018}, which have been attributed to the effects of diffusion. In general, diffusion causes surface abundances to decrease along the main sequence by up to $\sim$0.1 dex in certain elements \citep{Dotter+2017}. After the turnoff and up to the red giant branch, convection erases these effects, restoring the original abundances. The order of magnitude of diffusion is expected to be the largest in Na, Mg, Al, and Fe. We do see differences in Na and Mg (though for Mg the gradient is in the opposite direction), but we do not see a clear sign in Al and Fe.
\item This effect can also be due to systematics introduced by the analysis, for example, due to unidentified blends which are stronger for a certain temperature of the star \citep[see][for a review]{Jofre+2019}. Elements with few lines are probably more affected by this (e.g. Mg can be the case). Also, we see a difference between the abundances of \ion{Fe}{I} and \ion{Fe}{II}, the latter having a dependence in $\teff$ probably because of the fewer visible lines.
\end{enumerate}

In conclusion, bracket abundances are affected by different systematic effects that are difficult to disentangle, and that depends on spectral type and age in different ways. It is difficult to reach a conclusion on which effect is playing a significant role in each element. As seen in Sect.~\ref{sec:diff_abund}, this can be solved computing differential abundances.

Given the observed differences depending on the evolutionary state, and in order to give reference values of bracket abundances, we computed the median abundance and its MAD for different spectral types.
In Table~\ref{tab:cluster_abundances} we list the final bracket abundance values for F, G dwarfs and K giants of the three clusters. We remark that \ion{Fe}{I} and \ion{Fe}{II} are not in agreement in some cases, e.g. for F Dwarfs and K Giants in NGC~2632 differences are up to 0.07 dex. This is a direct consequence of the trend with temperature seen in \ion{Fe}{II} and not in the neutral case. On the contrary, in general, the two states of Ti seem to agree more among them.

\begin{table*}
\caption{Cluster average abundances (w.r.t. the Sun) and dispersions, weighted by the uncertainty. For each cluster we give a value of: F dwarfs, G dwarfs, and K giants. The number of stars in each group and element is indicated in parenthesis.}\label{tab:cluster_abundances}
\centering
\setlength\tabcolsep{1.5pt}
\small
\begin{tabular}{l|rrr|rr|rrr}
\hline
      & \multicolumn{3}{l}{Hyades} & \multicolumn{2}{l}{NGC 2632} & \multicolumn{3}{l}{Ruprecht 147} \\
$\mathrm{[X/H]}$ &             F Dwarfs &            G Dwarfs &            K Giants &             F Dwarfs &            K Giants &             F Dwarfs &           G Dwarfs &            K Giants \\
\hline
\ion{Na}{I} &   $0.09\pm0.04$ (23) &  $0.16\pm0.04$ (36) &   $0.46\pm0.02$ (3) &   $0.10\pm0.05$ (15) &   $0.49\pm0.06$ (3) &   $0.06\pm0.04$ (13) &  $0.10\pm0.01$ (4) &   $0.25\pm0.01$ (5) \\
\ion{Mg}{I} &   $0.05\pm0.04$ (23) &  $0.00\pm0.06$ (36) &   $0.06\pm0.03$ (3) &   $0.12\pm0.05$ (15) &   $0.06\pm0.03$ (3) &   $0.07\pm0.04$ (14) &  $0.04\pm0.04$ (4) &   $0.01\pm0.03$ (5) \\
\ion{Al}{I} &   $0.06\pm0.03$ (18) &  $0.14\pm0.02$ (36) &   $0.12\pm0.03$ (3) &   $0.08\pm0.06$ (15) &   $0.13\pm0.02$ (3) &  $-0.03\pm0.06$ (15) &  $0.08\pm0.03$ (4) &   $0.08\pm0.02$ (5) \\
\ion{Si}{I} &   $0.11\pm0.04$ (23) &  $0.15\pm0.03$ (24) &   $0.14\pm0.02$ (3) &   $0.13\pm0.03$ (15) &   $0.14\pm0.02$ (3) &   $0.03\pm0.03$ (14) &  $0.08\pm0.02$ (4) &   $0.11\pm0.02$ (5) \\
\ion{Ca}{I} &   $0.11\pm0.04$ (22) &  $0.17\pm0.03$ (36) &   $0.08\pm0.02$ (3) &   $0.12\pm0.05$ (15) &   $0.09\pm0.02$ (3) &   $0.03\pm0.03$ (14) &  $0.10\pm0.02$ (4) &  $-0.00\pm0.04$ (5) \\
\ion{Sc}{II}&   $0.03\pm0.06$ (23) &  $0.12\pm0.03$ (35) &   $0.10\pm0.02$ (3) &   $0.06\pm0.03$ (15) &   $0.11\pm0.02$ (3) &   $0.00\pm0.05$ (14) &  $0.05\pm0.03$ (4) &   $0.01\pm0.03$ (5) \\
\ion{Ti}{I} &   $0.06\pm0.03$ (18) &  $0.15\pm0.02$ (35) &   $0.08\pm0.03$ (3) &   $0.11\pm0.04$ (15) &   $0.07\pm0.03$ (3) &   $-0.01\pm0.01$ (7) &  $0.06\pm0.03$ (4) &  $-0.01\pm0.03$ (5) \\
\ion{Ti}{II}&   $0.09\pm0.04$ (21) &  $0.13\pm0.03$ (36) &   $0.09\pm0.01$ (3) &   $0.15\pm0.05$ (14) &   $0.05\pm0.02$ (3) &    $0.01\pm0.01$ (5) &  $0.07\pm0.02$ (3) &   $0.00\pm0.03$ (5) \\
\ion{V }{I}  &   $0.05\pm0.05$ (17) &  $0.19\pm0.03$ (36) &   $0.16\pm0.03$ (3) &   $0.07\pm0.04$ (11) &   $0.16\pm0.02$ (3) &   $-0.04\pm0.02$ (7) &  $0.09\pm0.02$ (4) &   $0.07\pm0.03$ (5) \\
\ion{Cr}{I} &   $0.10\pm0.04$ (20) &  $0.19\pm0.02$ (33) &   $0.13\pm0.02$ (3) &   $0.12\pm0.04$ (13) &   $0.15\pm0.02$ (2) &   $0.03\pm0.04$ (12) &  $0.12\pm0.01$ (3) &   $0.04\pm0.03$ (5) \\
\ion{Mn}{I} &   $0.04\pm0.06$ (21) &  $0.23\pm0.05$ (36) &   $0.05\pm0.02$ (3) &   $0.09\pm0.04$ (15) &   $0.04\pm0.03$ (3) &  $-0.01\pm0.05$ (14) &  $0.07\pm0.03$ (4) &  $-0.02\pm0.03$ (5) \\
\ion{Fe}{I} &   $0.10\pm0.05$ (21) &  $0.15\pm0.02$ (35) &   $0.10\pm0.02$ (3) &   $0.14\pm0.04$ (14) &   $0.10\pm0.03$ (3) &   $0.04\pm0.05$ (11) &  $0.11\pm0.02$ (3) &   $0.02\pm0.03$ (5) \\
\ion{Fe}{II}&   $0.15\pm0.07$ (23) &  $0.17\pm0.03$ (36) &   $0.06\pm0.02$ (3) &   $0.20\pm0.04$ (15) &   $0.03\pm0.02$ (3) &   $0.09\pm0.05$ (14) &  $0.11\pm0.03$ (4) &  $-0.02\pm0.05$ (5) \\
\ion{Co}{I} &   $0.08\pm0.03$ (18) &  $0.13\pm0.02$ (36) &   $0.15\pm0.03$ (3) &   $0.09\pm0.06$ (15) &   $0.15\pm0.03$ (3) &   $0.04\pm0.05$ (14) &  $0.07\pm0.03$ (4) &   $0.09\pm0.02$ (5) \\
\ion{Ni}{I} &   $0.11\pm0.04$ (16) &  $0.14\pm0.02$ (33) &   $0.10\pm0.03$ (3) &   $0.15\pm0.03$ (13) &   $0.10\pm0.02$ (3) &   $0.07\pm0.02$ (11) &  $0.08\pm0.03$ (4) &   $0.05\pm0.03$ (5) \\
\ion{Cu}{I} &   $0.08\pm0.06$ (22) &  $0.08\pm0.03$ (36) &   $0.01\pm0.02$ (3) &   $0.13\pm0.07$ (15) &   $0.02\pm0.03$ (3) &   $0.07\pm0.05$ (12) &  $0.08\pm0.03$ (4) &  $-0.05\pm0.03$ (5) \\
\ion{Y }{II} &   $0.14\pm0.08$ (23) &  $0.20\pm0.03$ (35) &   $0.37\pm0.02$ (3) &   $0.16\pm0.05$ (15) &   $0.41\pm0.04$ (3) &   $0.04\pm0.07$ (13) &  $0.07\pm0.09$ (4) &   $0.25\pm0.04$ (5) \\
\ion{Ba}{II}&   $0.17\pm0.08$ (23) &  $0.17\pm0.04$ (36) &   $0.21\pm0.02$ (3) &   $0.21\pm0.07$ (15) &   $0.21\pm0.01$ (3) &   $0.11\pm0.11$ (14) &  $0.14\pm0.06$ (4) &   $0.07\pm0.03$ (5) \\
\ion{La}{II}&                     &  $0.17\pm0.06$ (37) &  $-0.05\pm0.02$ (3) &                     &  $-0.05\pm0.02$ (3) &      -               &  $0.02\pm0.05$ (3) &  $-0.09\pm0.04$ (5) \\
\ion{Ce}{II}&                     &  $0.19\pm0.05$ (20) &   $0.20\pm0.03$ (3) &                     &   $0.19\pm0.03$ (3) &                     &  $0.26\pm0.00$ (1) &   $0.09\pm0.04$ (5) \\
\ion{Nd}{II}&  $-0.16\pm0.06$ (13) &  $0.14\pm0.06$ (27) &  $-0.20\pm0.02$ (3) &  $-0.14\pm0.03$ (12) &  $-0.21\pm0.02$ (3) &  $-0.11\pm0.17$ (13) &  $0.06\pm0.03$ (4) &  $-0.27\pm0.03$ (5) \\
\ion{Eu}{II}&                     &                    &   $0.17\pm0.02$ (3) &    $0.07$ (1) &   $0.12\pm0.02$ (3) &                     &                   &   $0.07\pm0.03$ (5) \\
\hline
\end{tabular}
\end{table*}

\subsection{\emph{Gaia} FGK Benchmark Stars}
The \gls{GBS} are a set of reference stars, which cover different regions of the HR diagram and a wide range in metallicity. For these stars, the effective temperature and surface gravity were determined independently from spectroscopy \citep{Heiter+2015,Hawkins+2016b}. Reference metallicities \citep{Jofre+2014,Hawkins+2016b} and abundances \citep{Jofre+2015} also exist. They are widely used in the community for cross-calibration and validation of pipelines and spectroscopic analyses. A summary description of their latest atmospheric parameters can be found in \citet{Jofre+2018}.

We used the sample of the \gls{GBS} high-quality spectra from the spectral library by \citet{Blanco+2014} to test the atmospheric parameters resulting from our pipeline. We also queried the archives for more spectra of these stars. We selected a subset of the whole sample of \gls{GBS} according to the parameter space covered by the cluster stars: those classified as FGK dwarfs (with $6500$~K $<\teff<4900$~K), FGK giants, and excluding metal-poor stars ($\mh<-1$). With this selection, we obtained 184 spectra of 16 \gls{GBS}. We processed these spectra using the same pipeline as the one used for the cluster stars.

\begin{figure}
\centering
\includegraphics[width=0.5\textwidth]{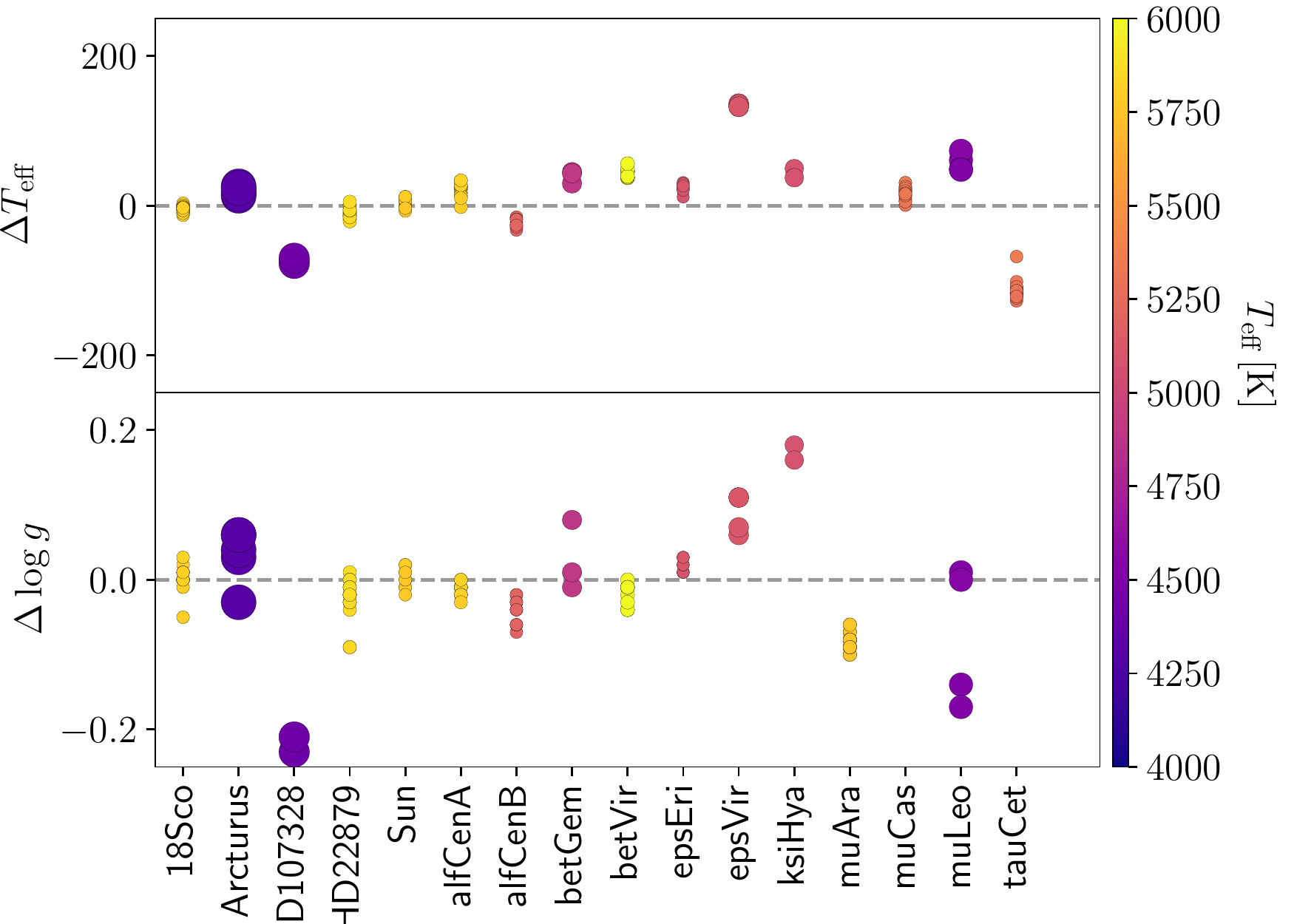}
\caption{Differences \citep[here -- reference values from][]{Jofre+2018} in $\teff$ and $\logg$ for the selection of the \gls{GBS}. The colors correspond to the temperature and the sizes are scaled with the inverse of the surface gravity (larger symbols mean giant stars). Vertically aligned symbols correspond to different spectra of the same star.}\label{fig:AP_GBS}
\end{figure}

We plot in Fig.~\ref{fig:AP_GBS} the obtained $\teff$ and $\logg$ compared with the reference \citep{Heiter+2015,Jofre+2018} for each spectrum.
We obtain a good agreement, with differences below 100 K and 0.1 dex in $\teff$ and $\logg$ respectively, with however larger differences for the giants HD~107328, $\epsilon$Vir and $\xi$Hya, especially for the surface gravity.
Remarkably seen in Fig.~\ref{fig:AP_GBS}, for a given star, the analysis of the spectra even from different instruments, return values in very good internal agreement up to 15 K in $\teff$ and 0.02 dex in $\logg$, except for two of the spectra of $\mu$Leo. A summary of the mean differences by spectral type (label "Group" indicated in the GBS reference table) is found in Table~\ref{tab:GBS_differences} where it is clear that we do not obtain any significant offset for a given spectral type.

\begin{table}
\caption{Comparison of the results of $\teff$ and $\logg$ for the GBS with respect to the reference ones (see text). Mean differences $\pm$ standard deviations are listed, together with the number of stars and spectra for each spectral type.}\label{tab:GBS_differences}
\centering
\begin{tabular}{lcccc}
\hline
Spectral & Num   & Num     & $\Delta\teff$ & $\Delta\logg$ \\
Type     & Stars & Spectra & (K)           & (dex)         \\
\hline
FGK giants    &   6       &    20       & $ 46\pm57$    & $0.01\pm0.11$  \\
G dwarfs      &   9       &   155       & $-17\pm49$    & $-0.03\pm0.03$ \\
K dwarfs      &   1       &     9       & $ 24\pm 5$    & $0.020\pm0.008$\\
\hline
All           &   16      &   184       & $-7\pm53$    & $-0.02\pm0.06$ \\
\hline
\end{tabular}
\end{table}

The abundance values of the different spectra of the analyzed stars are plotted in Fig.~\ref{fig:abu_GBS_spectr} for each analyzed element. The values of the median abundances per star and their uncertainties are listed in Tables~\ref{tab:GBS_abundances1} and \ref{tab:GBS_abundances2}. For most of the elements, the abundance dispersion (MAD) per star is small, on the order of 0.01 to 0.02 dex, reflecting the good agreement between lines of an element in the same star. The cases of Eu and Ce are the ones giving larger MAD, sometimes larger than 0.05 dex, this is possibly because their abundances are retrieved from the fit of very few weak lines in the spectra.

We did an external comparison using the reference values from \citet{Jofre+2015}. This study provides reference abundances of iron-peak and $\alpha$ elements for the whole sample of the \gls{GBS}. The mean values and standard deviations of the differences (this work -- reference) per element are indicated in Table~\ref{tab:GBS_meandif}. All the differences, which are always lower than 0.05 dex, are consistent with the obtained dispersions and the quoted uncertainties. An exhaustive comparison star by star is plotted in Fig~\ref{fig:abu_GBS}. 

\begin{table}
\caption{Mean (weighted by the uncertainty) element abundance difference obtained comparing this work with respect to the reference values of the \gls{GBS} \citet{Jofre+2015}.}\label{tab:GBS_meandif}
\centering
\begin{tabular}{lc}
\hline
Element & Mean difference \\
\hline
$\Delta\mathrm{[Fe/H]}$ & $-0.04\pm0.08$ \\
$\Delta\mathrm{[Ca/H]}$ & $-0.03\pm0.08$ \\
$\Delta\mathrm{[Co/H]}$ & $0.01\pm0.08$ \\
$\Delta\mathrm{[Cr/H]}$ & $-0.01\pm0.07$ \\
$\Delta\mathrm{[Mg/H]}$ & $-0.05\pm0.07$ \\
$\Delta\mathrm{[Mn/H]}$ & $0.02\pm0.08$ \\
$\Delta\mathrm{[Ni/H]}$ & $-0.02\pm0.08$ \\
$\Delta\mathrm{[Sc/H]}$ & $-0.04\pm0.08$ \\
$\Delta\mathrm{[Si/H]}$ & $-0.03\pm0.07$ \\
$\Delta\mathrm{[Ti/H]}$ & $-0.01\pm0.08$ \\
$\Delta\mathrm{[V/H]}$ & $0.01\pm0.08$ \\
\hline
\end{tabular}
\end{table}

\section{Differential chemical abundances}\label{sec:diff_abund}

Strictly line-by-line differential abundances are computed as a final step in our pipeline as explained in Sect.~\ref{sec:pipeline}.

Differential abundances have been computed in several previous works for solar twins or solar analogs using the Sun as a reference star \citep[e.g.][]{Melendez+2009b,TucciMaia+2016,Liu+2016}. This type of analysis provides high precision abundances, erasing most of the effects which blur typical chemical abundance procedures like unaccounted blends, effects of stellar evolution and poor atomic line characterization. This technique has also been applied to other stellar types, selecting a reference star which is as close as possible to the analyzed stars in terms of stellar parameters \citep{Reggiani+2017,Hawkins+2016b,Jofre+2015}. In this case, the results tell us how much the abundances of the analyzed stars differ from those of the reference star, which is no longer necessarily the Sun. This is a good strategy to perform chemical tagging experiments.

\begin{figure}
\centering
\includegraphics[width=0.5\textwidth]{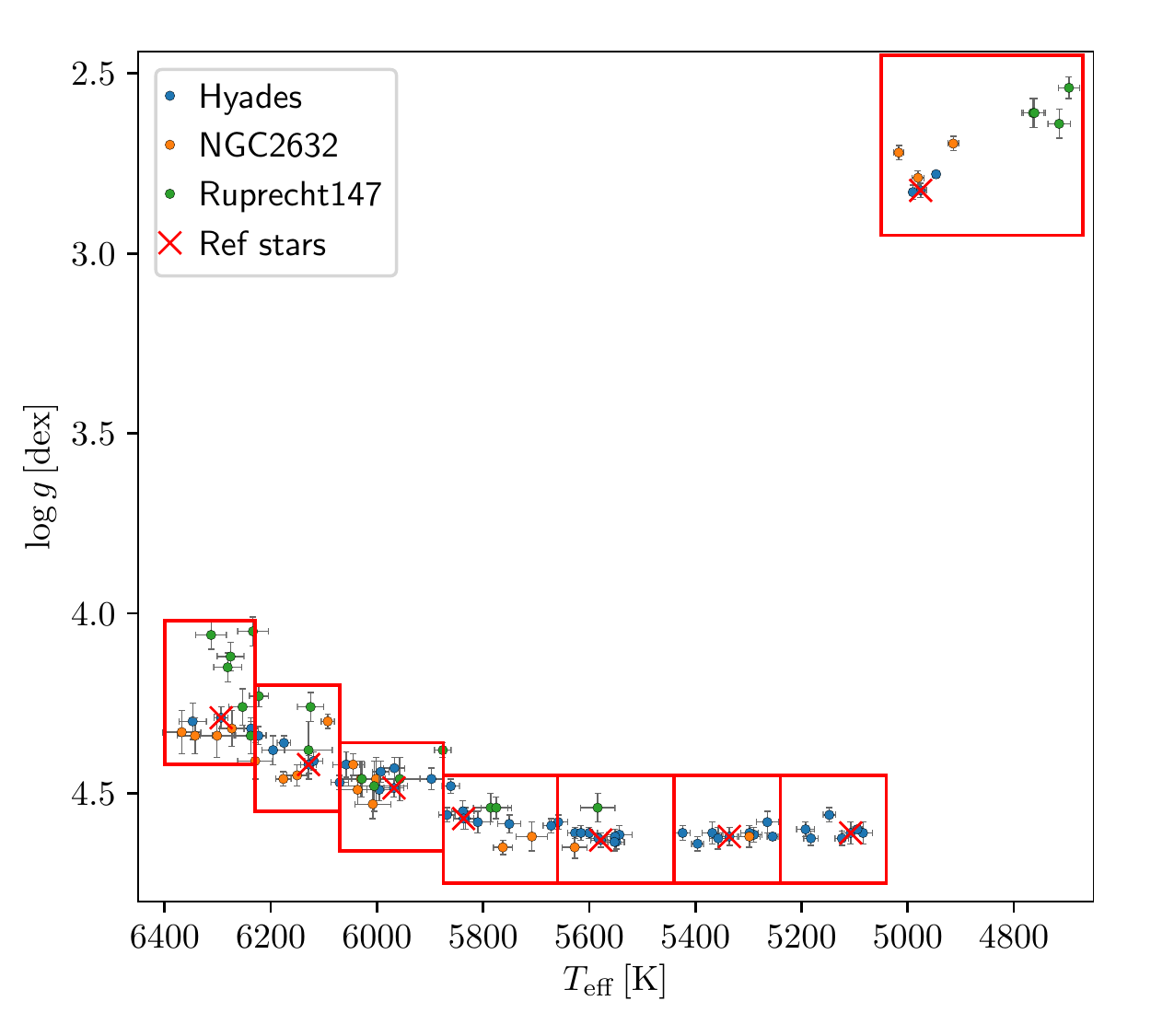}
\caption{HR diagram of the grouped stars in the three clusters. Colors represent each cluster as in Fig.~\ref{fig:AP}. Red boxes indicate the $\teff$ and $\logg$ limits of the groups used for the differential analysis. Red crosses mark the chosen reference stars.}\label{fig:groups}
\end{figure}

In the case of the stars in the OCs analyzed here, we have a large spread in atmospheric parameters, and so performing a differential analysis is challenging. Therefore, we developed a strategy to perform a differential analysis not only for solar twins but using stars at any evolutionary stage. We made 8 groups of stars which differ among them by less than $\sim$200 K, and $\sim$0.3 dex in $\teff$ and $\logg$, respectively, to be analyzed together as twins. For the upper main sequence and the giants the limits in $\logg$ and $\teff$ were relaxed to 0.45 dex, in order to include in the differential analysis the warmest stars in the upper main sequence with the rest of the stars of the same temperature, and the giants of Ruprecht~147 together with the giants from the other two clusters. The groups have between 4 and 16 stars in the three OCs and follow the main sequence and the giants. One star in each group was selected to be used as the reference to compute the differential abundances. See Fig.~\ref{fig:groups}. Among the member stars in the three OCs, we were able to group 92 stars, leaving out those which fall out of the group limits in the $\teff-\logg$ plane.

The resulting abundance value of the element $X$, designated as $\delta X$, was computed as the mean of the abundance difference with respect to the reference for each line:

\begin{equation}
\delta X = \frac{1}{N_{lines}}\sum_{i=1}^{N_{lines}} \left(A_{X_i}-A_{X_i,REF}\right)    
\end{equation}

The reference stars selected for each group are different from each other. This could have consequences if one of the chosen stars has a chemical peculiarity. In this case, the abundance scale for that group would be different from the others. For our experiment, the reference stars were chosen to be stars from the Hyades cluster that fall approximately in the middle of the $\teff$ and [Fe/H] range of each group. We have checked for previous information of these stars to be sure they have not been identified as high rotators or peculiar stars. We list the chosen reference stars with their atmospheric parameters in Table~\ref{tab:refstars}.

\begin{table}
\caption{Reference stars used to compute differential abundances in each defined group. We indicate the \emph{Gaia} DR2 source ID, Hipparcos ID, the computed atmospheric parameters, iron bracket abundance and the number of analyzed spectra. For stars with more than one spectra we indicate the mean values and standard deviations of all determinations.\label{tab:refstars}}
\centering
\def\arraystretch{1.5}
\setlength\tabcolsep{2pt}
\begin{tabular}{lcccc}
\hline
Star                & $\teff$ (K) & $\logg$ (dex) & [Fe/H] (dex) & N\\
\hline
$^{\rm Gaia\,DR2\,3300934223858467072}_{\rm HIP\,19796}$  & $6286\pm17$ & $4.30\pm0.04$ & $0.13\pm0.06$ & 3 \\
$^{\rm Gaia\,DR2\,48203487411427456  }_{\rm HIP\,20237}$  & $6126\pm19$ & $4.41\pm0.04$ & $0.11\pm0.06$ & 3 \\
$^{\rm Gaia\,DR2\,3314109916508904064}_{\rm HIP\,20899}$  & $5957\pm32$ & $4.49\pm0.03$ & $0.12\pm0.05$ & 4 \\
$^{\rm Gaia\,DR2\,3313689422030650496}_{\rm HIP\,20741}$  & $5834\pm19$ & $4.57\pm0.04$ & $0.17\pm0.05$ & 5 \\
$^{\rm Gaia\,DR2\,144171233106399104 }_{\rm HIP\,21099}$  & $5582\pm23$ & $4.63\pm0.02$ & $0.14\pm0.04$ & 3 \\
$^{\rm Gaia\,DR2\,3406823245223942528}_{\rm HIP\,22380}$  & $5351\pm20$ & $4.61\pm0.04$ & $0.14\pm0.06$ & 1 \\
$^{\rm Gaia\,DR2\,64266768177592448  }_{\rm HIP\,16908}$  & $5107\pm16$ & $4.61\pm0.03$ & $0.16\pm0.05$ & 1 \\
$^{\rm Gaia\,DR2\,3312052249216467328}_{\rm HIP\,20205}$  & $4975\pm12$ & $2.83\pm0.03$ & $0.10\pm0.05$ & 2 \\
\hline
\end{tabular}
\end{table}
      
\begin{figure*}
\centering
\includegraphics[width=\textwidth]{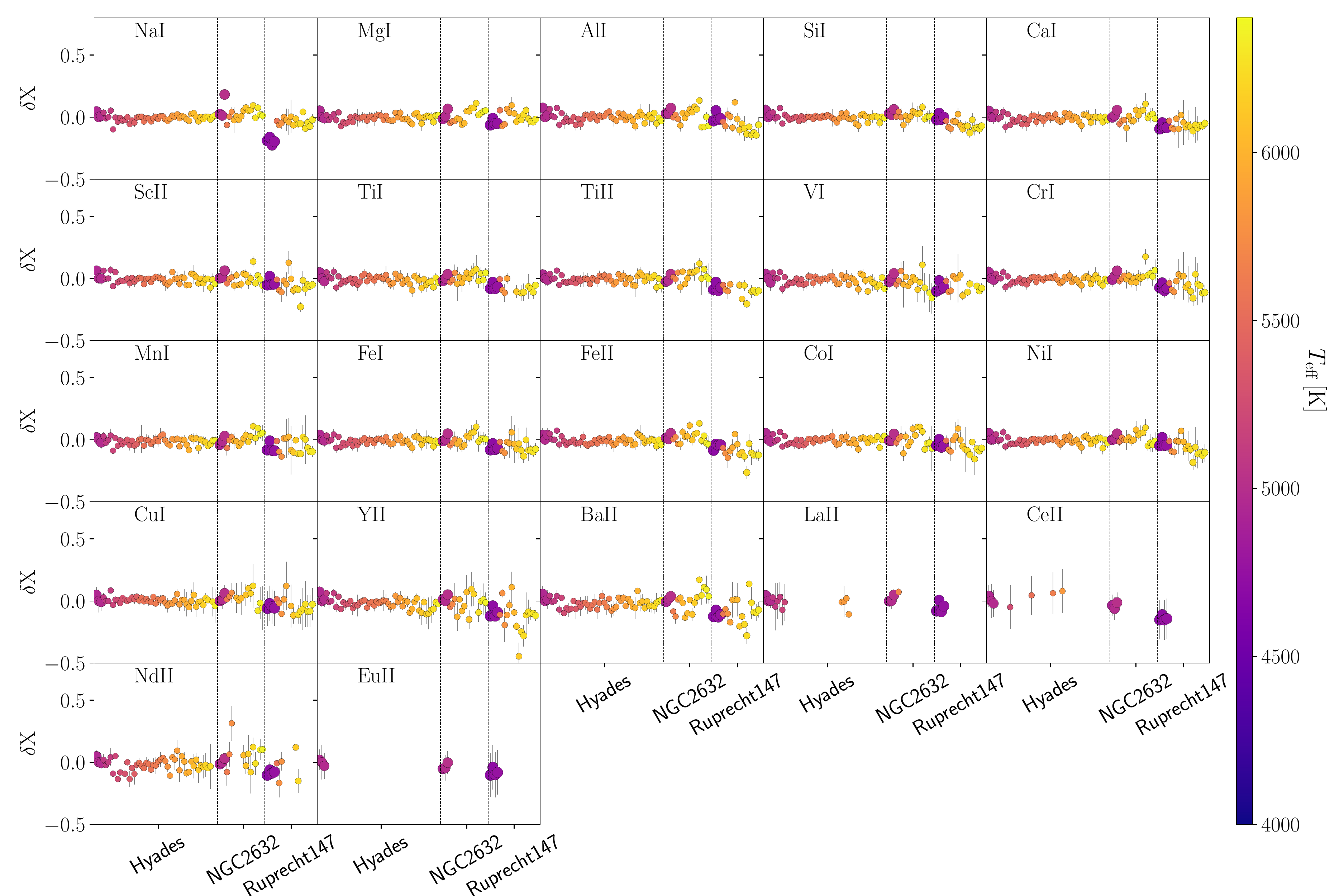}
\caption{Differential abundances as a function of effective temperature computed for the analyzed OCs. Colors and sizes represent the same as in Fig.~\ref{fig:absolute_abus}.}\label{fig:differential_abus}
\end{figure*}

The resulting differential abundances for all chemical species are plotted for the three OCs in Fig.~\ref{fig:differential_abus} as a function of $\teff$. We have discarded the two stars in the tidal tails analyzed in Sec.~\ref{sec:tails} for being not members according to their chemical abundances.

Compared with the equivalent figure with bracket abundances (Fig.~\ref{fig:absolute_abus}) the dependence with temperature has erased for most of the elements.

\subsection{Precision in differential abundances}\label{sec:precision}
To assess our precision, we independently analyzed nine spectra of the same star, obtained with five different instruments. The chosen star is a dwarf from the Hyades, Gaia DR2 3313689422030650496. It was selected because it has spectra with different \gls{SNR} values, representative of the range that we have for the whole sample.

\begin{figure}
\centering
\includegraphics[width=0.5\textwidth]{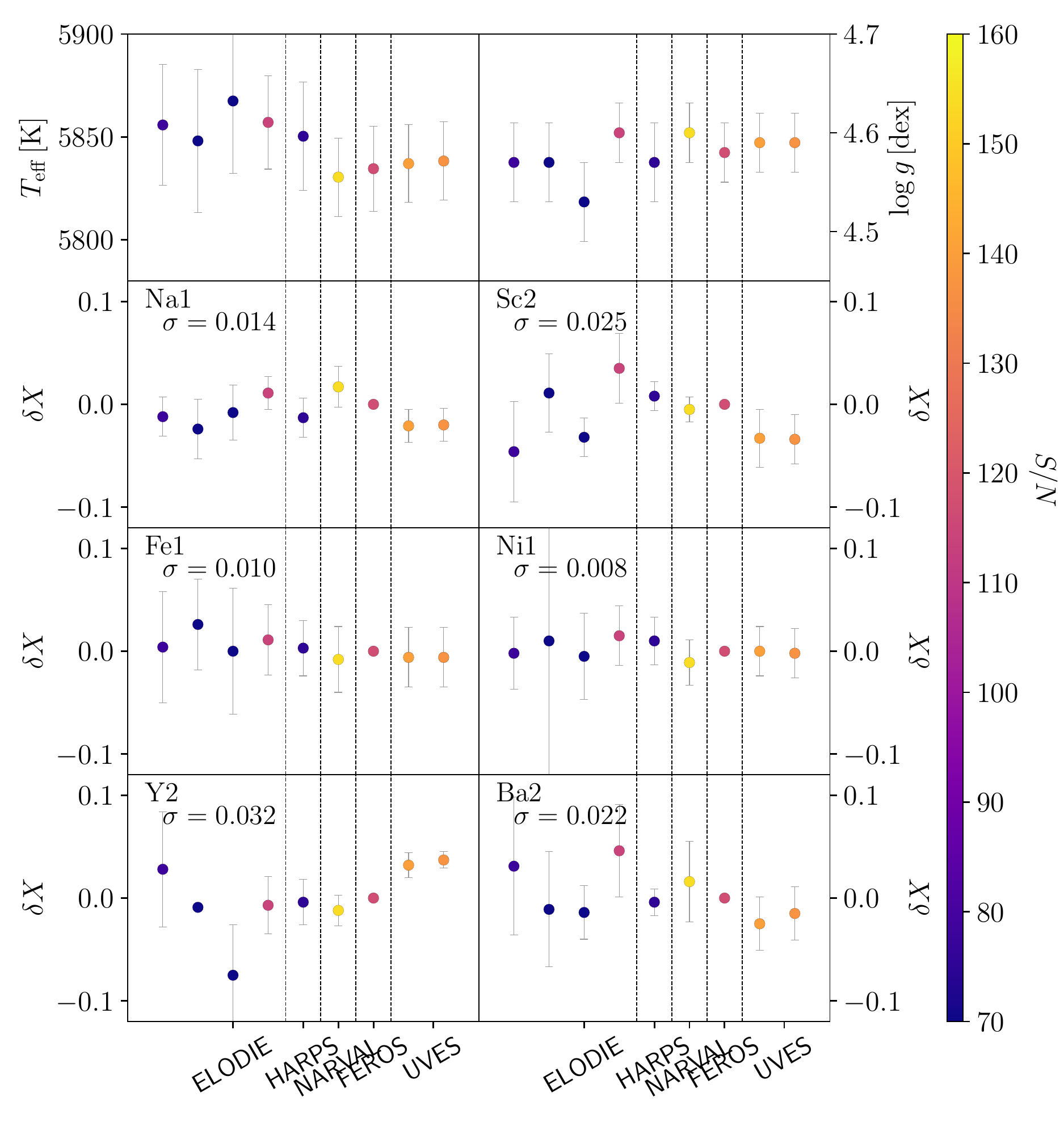}
\caption{First row: atmospheric parameters ($\teff$ and $\logg$) retrieved from the analysis of the different spectra of the star Gaia DR2 3313689422030650496. Last three rows: differential abundances of several chemical species for all spectra. The color code represents the \gls{SNR} of the spectra.}\label{fig:differential_APabus_repstar}
\end{figure}

The resulting atmospheric parameters give a mean $\teff$ of 5847 K (standard deviation of 11 K, and a mean uncertainty of 25 K), and the mean $\logg$ is 4.58 dex (standard deviation of 0.02 dex, and mean of the quoted uncertainty of 0.03 dex).
For the chemical abundance computation, we used the FEROS spectrum as a reference, and we computed line-by-line differential abundances using the same analysis pipeline as for the bulk of the OC stars.
In Fig~\ref{fig:differential_APabus_repstar} we plot the resulting $\teff$, $\logg$, and abundances of six representative species for all the spectra of this star. In Table~\ref{tab:repstar_stds} we list the dispersions in differential abundances found for all elements and the mean of the quoted uncertainties.

We do not obtain any clear systematics between instruments in the atmospheric parameters or the differential abundances. In general, the dispersions in abundances are on the order of 0.01 to 0.02 dex. The mean quoted uncertainties tend to be larger (0.03 dex) compared with the dispersions, which indicates that we might have slightly overestimated uncertainties. The largest variations are seen for spectra which have \glspl{SNR} at the lower end, and for elements which present intrinsic difficulties to our pipeline for measuring abundances because of few or weak lines, e.g. Y.

\begin{table}
\caption{Dispersions and mean quoted uncertainties of the differential abundances found with the different spectra of the dwarf star Gaia DR2 3313689422030650496. For elements without an uncertainty the abundance comes from a single line. No \ion{La}{II} and \ion{Nd}{II} abundances are computed for this star.}\label{tab:repstar_stds}
\centering
\begin{tabular}{lcc}
\hline
Element & Abundance  & Mean  \\
        & dispersion & uncertainty \\
\hline
 $\ion{Na}{1}$ & $0.016$ & $0.019$ \\
$\ion{Mg}{1}$ & $0.015$ & $0.007$ \\
$\ion{Al}{1}$ & $0.015$ & $0.010$ \\
$\ion{Si}{1}$ & $0.008$ & $0.027$ \\
$\ion{Ca}{1}$ & $0.013$ & $0.025$ \\
$\ion{Sc}{2}$ & $0.023$ & $0.018$ \\
$\ion{Ti}{1}$ & $0.017$ & $0.028$ \\
$\ion{Ti}{2}$ & $0.018$ & $0.034$ \\
$\ion{V}{1}$  & $0.015$ & $0.041$ \\
$\ion{Cr}{1}$ & $0.009$ & $0.039$ \\
$\ion{Mn}{1}$ & $0.025$ & $0.020$ \\
$\ion{Fe}{1}$ & $0.011$ & $0.033$ \\
$\ion{Fe}{2}$ & $0.022$ & $0.049$ \\
$\ion{Co}{1}$ & $0.013$ & $0.032$ \\
$\ion{Ni}{1}$ & $0.007$ & $0.037$ \\
$\ion{Cu}{1}$ & $0.026$ & $0.008$ \\
$\ion{Y}{2}$  & $0.032$ & $0.021$ \\
$\ion{Ba}{2}$ & $0.023$ & $0.029$ \\
$\ion{Ce}{2}$ & $0.057$ & - \\
$\ion{Eu}{2}$ & $0.076$ & - \\
\hline
\end{tabular}
\end{table}

\subsection{Systematic uncertainties due to errors in atmospheric parameters}\label{sec:uncertaintiesAP}

We quantified the errors due to uncertainties in the stellar parameters in the same way as \citet{Jofre+2015}. We computed the differential abundances in using the same process as explained above but changing the value of the stellar parameters according to their uncertainties. The procedure was repeated eight times, adding and subtracting the error in each stellar parameter: $\teff$, $\logg$, $\vmic$, and $\mathrm{[Fe/H]}$. For each parameter, the difference between the two values of the resulting abundance (adding and subtracting) is considered as the uncertainty: $\Delta\teff$, $\Delta\logg$, $\Delta\vmic$, and $\Delta\mathrm{[Fe/H]}$. Then we consider the total uncertainty would be

\begin{equation}
\Delta = \sqrt{\Delta\teff^2 + \Delta\logg^2 + \Delta\vmic^2 + \Delta\mathrm{[Fe/H]}^2} \label{eq:uncert}
\end{equation}

if the four parameters are statistically independent, which is not the case. A full covariance matrix should be computed, and then the total uncertainty would be smaller. However, this is a simpler and more conservative way of taking into account these uncertainties. See a more extended discussion in \citet{Jofre+2019}.

\begin{figure}
\centering
\includegraphics[width=0.5\textwidth]{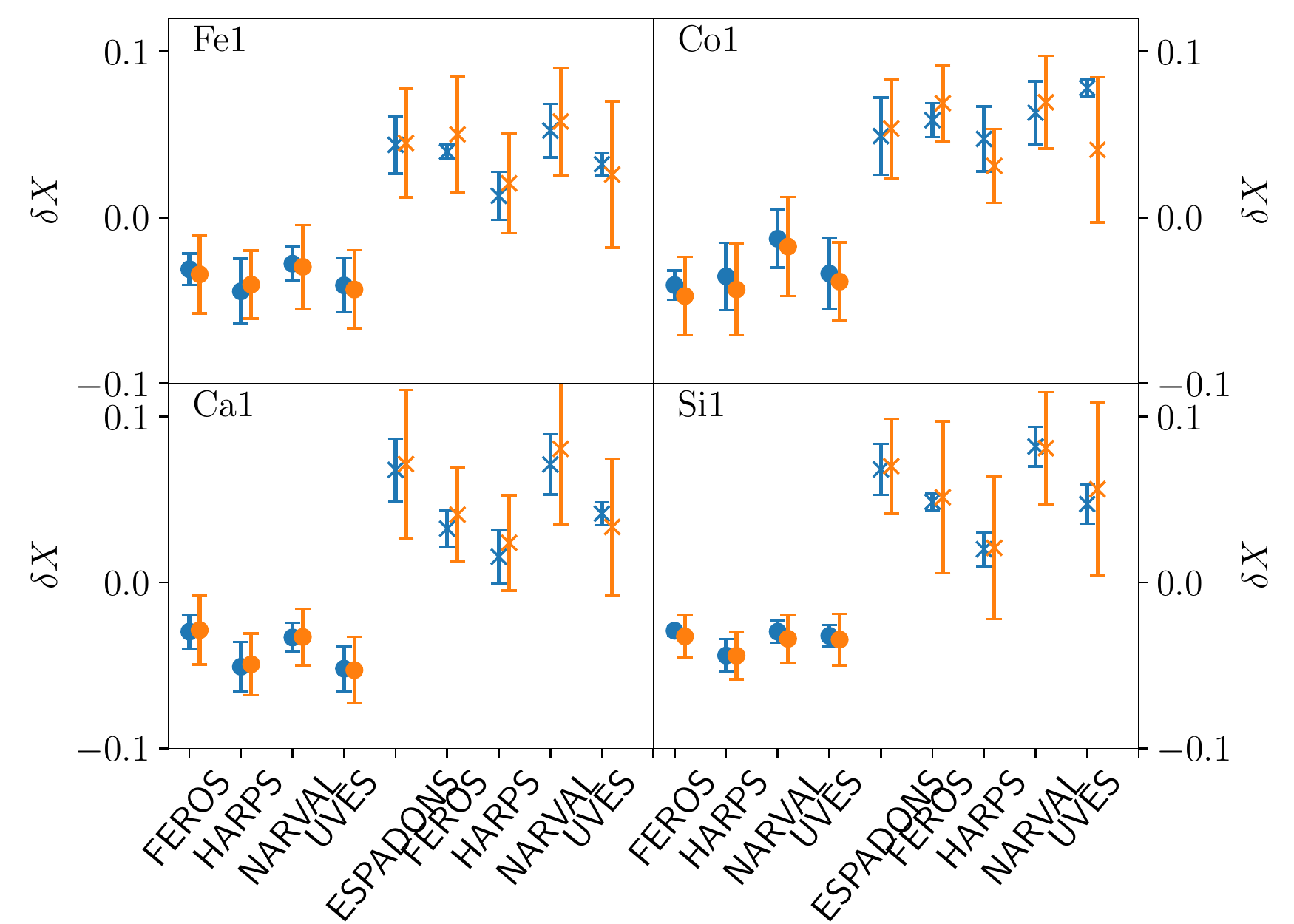}
\caption{Differential abundance results for different spectra of two stars, a dwarf (filled circles) and a giant (crosses). The values and uncertainties using the atmospheric parameters retrieved in Sec.~\ref{sec:APs} are in orange, and the values derived with the uncertainties in stellar parameters are in blue (see text).}\label{fig:uncert_AP}
\end{figure}

We have done this procedure for two representative stars observed with different instruments, a dwarf and a giant. The results for four chemical species are plotted in Fig.\ref{fig:uncert_AP}. We represent in blue the mean value of the abundances that resulted to change the atmospheric parameters by their uncertainties, and the error bar represents the $\Delta$ from Eq.~\ref{eq:uncert}. For all chemical species, the quoted uncertainties in the differential abundances are larger than the change in the abundance due to atmospheric parameters. This shows that somehow these uncertainties are already representative of the effect of the uncertainties due to the atmospheric parameters.

\subsection{On the homogeneity of OCs}

The level of homogeneity of OCs has been investigated in several recent studies, using differential analysis of solar twins. \citet[][hereafter L16]{Liu+2016} studied 16 stars in the Hyades, \citet{Liu+2016b} studied two stars in M~67, and \citet{Spina+2018} analyzed five stars in the Pleiades.
In the most extensive one, L16 computed abundances of 19 chemical species using high \gls{SNR} spectra of solar-type stars ($\teff\sim5600-6300$ K, $\logg\sim4.47-4.58$). They found total abundance variations of around $0.1$ dex, and an abundance scatter from 0.020 to 0.045 dex (depending on the chemical species), in general, larger than the quoted uncertainties by a factor of $\sim1.5$. Moreover, they found statistically significant correlations among the abundances of pairs of elements. They interpreted this as a signature of a genuine abundance scatter in the Hyades since it would mean that there is a difference in the overall metallicity between different stars. Otherwise, if differences appeared only for certain elements or for certain stars (i.e. scatter plot without correlations), this could be caused either by random errors or by certain processes in the stellar atmospheres depending on stellar parameters.

In this section, we investigate the homogeneity of the analyzed OCs. 
For the case of the Hyades, we have a larger sample than L16, 51 bonafide member stars for which we could compute differential abundances. We cover a range in $\teff$ of 4900$-$6300 K, and in $\logg$ of 2.8$-$4.65 dex. For NGC~2632 and Ruprecht~147 we obtain differential abundances of 19 and 20 members, respectively.

We show in Table~\ref{tab:cluster_difabus} the dispersion and amplitude of the differential abundances, and the mean of the quoted uncertainties, for each chemical species and each of the three clusters. The amplitudes of all elements are around 0.15 dex for the Hyades, and 0.2 dex in the other two clusters. The abundance dispersions for the Hyades are typically of the order of 0.02-0.03 dex. In general, these are not larger than the quoted uncertainties, however, they are larger than the dispersions among the results of different spectra of the same star found in Sec.~\ref{sec:precision} (lower than 0.02 dex, in general). We consider that this fact shows that our uncertainties are slightly overestimated.
For comparison, we also include the same computation for the subsample of 16 stars analyzed by L16, for the elements in common. 
In this smaller range of atmospheric parameters corresponding to solar analogs, we obtain slightly lower abundance dispersions similar to those of L16, ranging from 0.017 to 0.035 dex.
For NGC~2632 and Ruprecht~147 both the quoted uncertainties and the found dispersions are larger than for the Hyades in most of the chemical species. We cannot clearly conclude that the clusters are inhomogeneous only based on the comparison of the uncertainties with the abundance dispersions.

We obtain strong correlations among different abundance pairs in the three clusters. We performed linear fits to each pair of element abundances using a Bayesian outlier detector with a Markov Chain Monte Carlo (MCMC), as explained in detail in \citet{Hogg+2010}. In brief, the methodology computes a linear regression with an objective datapoint rejection, which models the outlier distribution. The method infers at once the parameters of the linear fit together with the mean and variance of the distribution of outliers, and the number of outliers. The model is run through $50\,000$ MCMC realizations, taking the maximum of the posterior distribution and the standard deviation as the best values of the slope, intercept and their uncertainties. We do not attempt to perform the fits when the number of stars with abundances is smaller than 15, or the uncertainties in the abundances are greater than 0.05 dex.

\begin{figure}
\centering
\includegraphics[width=0.5\textwidth]{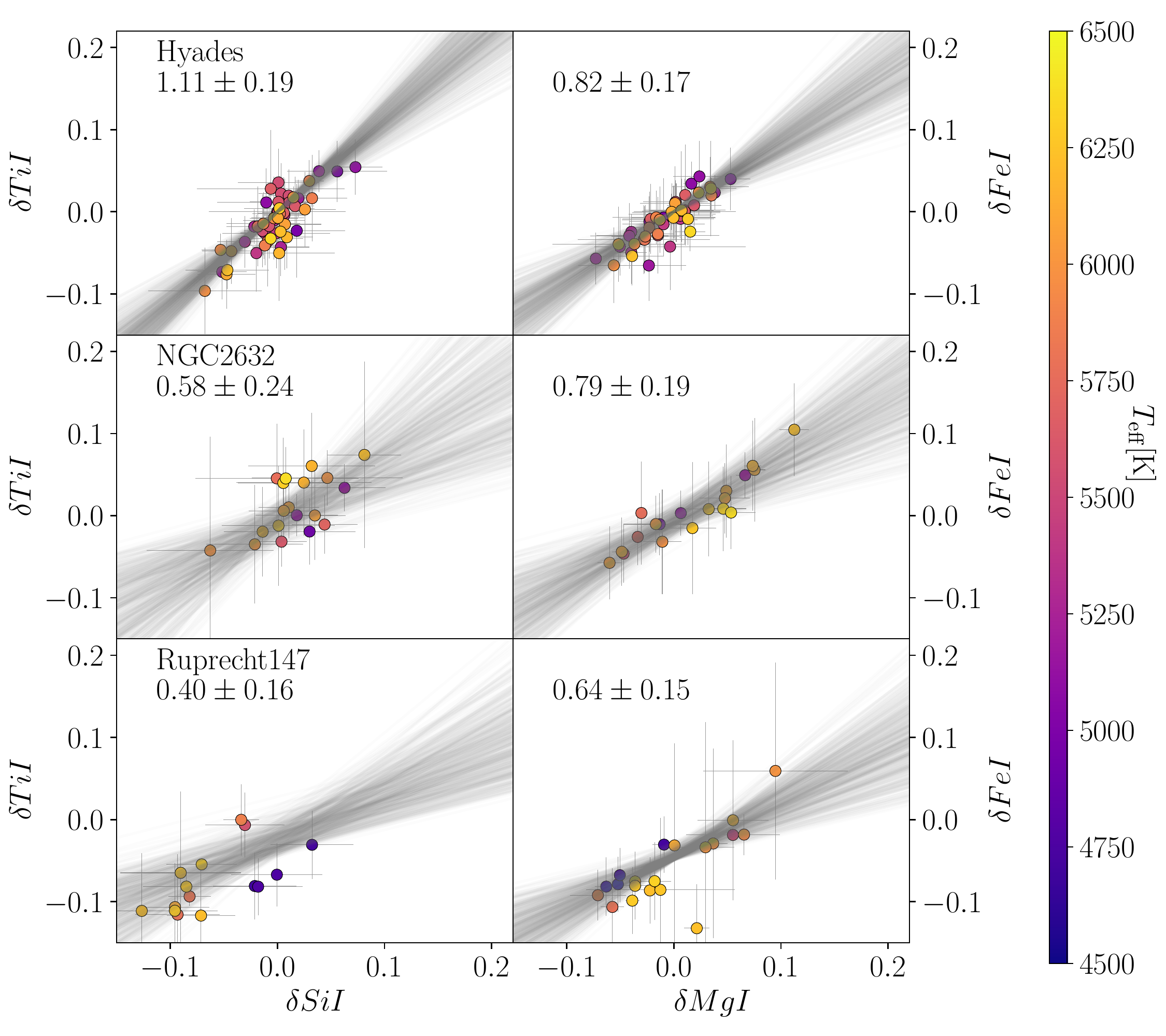}
\caption{Left: $\delta \mathrm{Ti}$ vs $\delta \mathrm{Si}$, right: $\delta \mathrm{Mg}$ vs $\delta \mathrm{Fe}$, rows correspond to the three clusters. Color represents effective temperature. Translucid lines are the resulting linear fit according to the obtained posterior distribution in the intercept and the slope. The slope and its uncertainty is indicated in each panel.}\label{fig:XHXH_OCs}
\end{figure}

We plot in Fig.~\ref{fig:XHXH_OCs} some examples of differential element abundance pairs $\delta X_1$ vs the $\delta X_2$. The full matrix of differential abundances for the three clusters are found in Figs.~\ref{fig:XHXH_Hyades},\ref{fig:XHXH_NGC2632}, and \ref{fig:XHXH_Rup147}. We also over plot the obtained fits.
For the Hyades, we obtain statistically significant correlations (larger than $3\sigma$) of all pairs of abundances that we analyze. Some elements with larger intrinsic uncertainties (e.g. Ba, Nd, Cu) tend to present less significant correlations.
For the other two clusters, we have a smaller sample of stars selected as most probable members, and larger uncertainties in the abundance measures. The less significant fit for with NGC~2632 is the one of \ion{Na}{I} vs \ion{Mn}{I}, which has a posterior with two peaks. For Ruprecht~147 two fits involving \ion{Si}{I} have lower significances than $3\sigma$. In Fig.~\ref{fig:XHXH_OCs} we see no dependence of the differential abundance pairs with respect to $\teff$, showing that all types of stars correlate in the same way.

For the Hyades we have a large amount of slopes retrieved, so we compute the compatibility among all of them within the quoted uncertainties:
(i) The fits involving \ion{Cu}{I} and \ion{Ba}{II} which have non-Gaussian posterior distributions tend to be not compatible with the others. However, in these cases, uncertainties are not well represented by the standard deviation. (ii) In the case of \ion{Nd}{II}, all slopes are larger than $3\sigma$ compared to most of the other cases. We attribute this to the larger uncertainties of this element in comparison with the others which makes the dispersion in its dimension increase, and so increase the value of the slope. (iii) We find compatible slopes for all other pairs of elements.
This indicates that the differences in abundance star-by-star are due to a zero point difference in the overall metallicity among the stars in the cluster, with no variation depending on the chemical element.

\begin{table*}
\caption{Left: Dispersion ($\sigma X$) and full amplitude range of the abundances ($A_{X}$, i.e. difference between highest and lowest value), and mean of the quoted uncertainties for each cluster. Right: Comparison with L16 results for the elements in common: dispersions obtained using the subsample of stars of L16 ($\sigma X_{\mathrm{here}}$), and their values ($\sigma X_{\mathrm{L16}}$).}\label{tab:cluster_difabus}
\centering
\setlength\tabcolsep{4.3pt}
\begin{tabular}{l|rrr|rrr|rrr}
\hline
Cluster & \multicolumn{3}{c}{Hyades} & \multicolumn{3}{c}{NGC 2632} & \multicolumn{3}{c}{Ruprecht 147} \\
\hline
        & $\sigma X$ & $A_{X}$ & Unc. & $\sigma X$  & $A_{X}$ & Unc. & $\sigma X$  & $A_{X}$ & Unc. \\
\hline
\ion{Na}{1} & 0.025 & 0.151 & 0.021 & 0.052 & 0.245 & 0.024 & 0.070 & 0.268 & 0.035 \\
\ion{Mg}{1} & 0.026 & 0.126 & 0.026 & 0.048 & 0.172 & 0.013 & 0.047 & 0.166 & 0.019 \\
\ion{Al}{1} & 0.033 & 0.149 & 0.029 & 0.057 & 0.214 & 0.021 & 0.067 & 0.264 & 0.038 \\
\ion{Si}{1} & 0.026 & 0.141 & 0.028 & 0.031 & 0.144 & 0.044 & 0.042 & 0.159 & 0.039 \\
\ion{Ca}{1} & 0.029 & 0.129 & 0.032 & 0.044 & 0.193 & 0.047 & 0.031 & 0.128 & 0.070 \\
\ion{Sc}{2} & 0.032 & 0.141 & 0.034 & 0.047 & 0.195 & 0.041 & 0.065 & 0.355 & 0.046 \\
\ion{Ti}{1} & 0.033 & 0.151 & 0.035 & 0.034 & 0.116 & 0.056 & 0.037 & 0.117 & 0.046 \\
\ion{Ti}{2} & 0.029 & 0.135 & 0.035 & 0.044 & 0.190 & 0.057 & 0.043 & 0.175 & 0.043 \\
\ion{V}{1} & 0.035 & 0.155 & 0.033 & 0.062 & 0.268 & 0.072 & 0.049 & 0.166 & 0.045 \\
\ion{Cr}{1} & 0.026 & 0.121 & 0.038 & 0.045 & 0.207 & 0.058 & 0.053 & 0.192 & 0.076 \\
\ion{Mn}{1} & 0.028 & 0.130 & 0.036 & 0.044 & 0.173 & 0.036 & 0.053 & 0.160 & 0.058 \\
\ion{Fe}{1} & 0.026 & 0.108 & 0.035 & 0.040 & 0.162 & 0.050 & 0.043 & 0.192 & 0.062 \\
\ion{Fe}{2} & 0.028 & 0.125 & 0.035 & 0.045 & 0.170 & 0.039 & 0.061 & 0.309 & 0.046 \\
\ion{Co}{1} & 0.029 & 0.140 & 0.027 & 0.059 & 0.213 & 0.043 & 0.055 & 0.246 & 0.051 \\
\ion{Ni}{1} & 0.027 & 0.128 & 0.031 & 0.039 & 0.147 & 0.051 & 0.056 & 0.254 & 0.068 \\
\ion{Cu}{1} & 0.026 & 0.134 & 0.045 & 0.050 & 0.201 & 0.099 & 0.055 & 0.237 & 0.114 \\
\ion{Y}{2} & 0.040 & 0.180 & 0.042 & 0.065 & 0.245 & 0.069 & 0.115 & 0.557 & 0.056 \\
\ion{Ba}{2} & 0.038 & 0.179 & 0.044 & 0.079 & 0.305 & 0.033 & 0.089 & 0.417 & 0.056 \\
\ion{La}{2} & 0.042 & 0.154 & 0.098 & 0.031 & 0.074 & 0.049 & 0.037 & 0.092 & 0.044 \\
\ion{Ce}{2} & 0.044 & 0.129 & 0.145 & 0.021 & 0.051 & 0.087 & 0.017 & 0.042 & 0.162 \\
\ion{Nd}{2} & 0.050 & 0.229 & 0.061 & 0.097 & 0.394 & 0.086 & 0.083 & 0.287 & 0.075 \\
\ion{Eu}{2} & 0.020 & 0.049 & 0.110 & 0.022 & 0.049 & 0.095 & 0.025 & 0.063 & 0.179 \\
\hline
\end{tabular}
\setlength\tabcolsep{4.3pt}
\begin{tabular}{l|rrr|rrr|rrr}
\hline
Element & $\sigma X_{\mathrm{L16}}$ & $\sigma X_{\mathrm{here}}$ \\
\hline
Na 1 & 0.021 & 0.017 \\
Mg 1 & 0.035 & 0.024 \\
Al 1 & 0.046 & 0.029 \\
Si 1 & 0.023 & 0.025 \\
Ca 1 & 0.023 & 0.023 \\
Ti 1 & 0.029 & 0.029 \\
Ti 2 & 0.032 & 0.028 \\
V 1  & 0.026 & 0.037 \\
Cr 1 & 0.026 & 0.021 \\
Mn 1 & 0.026 & 0.031 \\
Fe 1 & 0.023 & 0.024 \\
Co 1 & 0.030 & 0.024 \\
Ni 1 & 0.028 & 0.023 \\
Cu 1 & 0.036 & 0.025 \\
Ba 2 & 0.031 & 0.037 \\
\hline
\end{tabular}
\end{table*}

\subsubsection{Comparison with synthetic data}

To better interpret our results we generated a set of simulated spectra of stars for two cases: a chemically homogeneous cluster and an inhomogeneous one. In this way, we can see what we should expect in the case of chemical inhomogeneity and compare it with the results obtained for the observed spectra.

For both the homogeneous and the inhomogeneous cases we generated spectra of a set of stars with the same atmospheric parameters. We used parameters corresponding to the ones of two of the groups observed in the Hyades. That is, nine stars with $\teff=6000-5850$ K and $\logg=4.4-4.6$ dex, and six stars with $\teff=5850-5750$ K, $\logg=4.4-4.6$ dex. For the homogeneous case, we used a model metallicity of $\mathrm{[M/H]} = 0$ dex for all stars, whereas for the inhomogeneous case we sampled a normal distribution with a mean metallicity of 0 dex and a standard deviation of 0.03 dex. We added Poissonian noise to the spectra to degrade the \gls{SNR} to between 100 and 200. We analyzed both sets of spectra using the same procedure as for the real data.

Even if we introduce noise to the synthetic data, the experiment will not be comparable to the analysis of real data. For example, the effects of the continuum normalization and bad/missing atomic data are not taken into account and can contribute to the dispersion of a synthetic homogeneous cluster. To be more realistic, we added some possible sources of systematic uncertainties. To generate the spectra we used the radiative transfer code SYNTHE \citep{Kurucz1993} and the ATLAS9 \citep{Castelli+2004}atmospheric models, instead of SPECTRUM \citep{Gray+1994} and MARCS \citep{Gustafsson2008}, which are used in the analysis pipeline. We slightly altered the line shapes varying the macroturbulence parameter. In the analysis $\varv_{mac}$ is determined using the empirical relation given in \citet{BlancoCuaresma+2014}, which depends on $\teff$, $\logg$, and $\mathrm{[Fe/H]}$. To generate the spectra we added a random term ($\pm0.15$ \kms) to the estimated value.

We compared the recovered effective temperature and surface gravity with the input ones in the two cases, and we obtain almost the same small offsets on average (recovered $-$ input): $\Delta\teff \sim -16$ K, $\Delta\logg \sim -0.03$. For $\mathrm{[M/H]}$ the inhomogeneous case gives a larger offset and dispersion ($\Delta \mathrm{[M/H]}=-0.023 \pm 0.018$; standard deviation) than the homogeneous one ($-0.009 \pm 0.005$).

\begin{figure}
\centering
\includegraphics[width=0.5\textwidth]{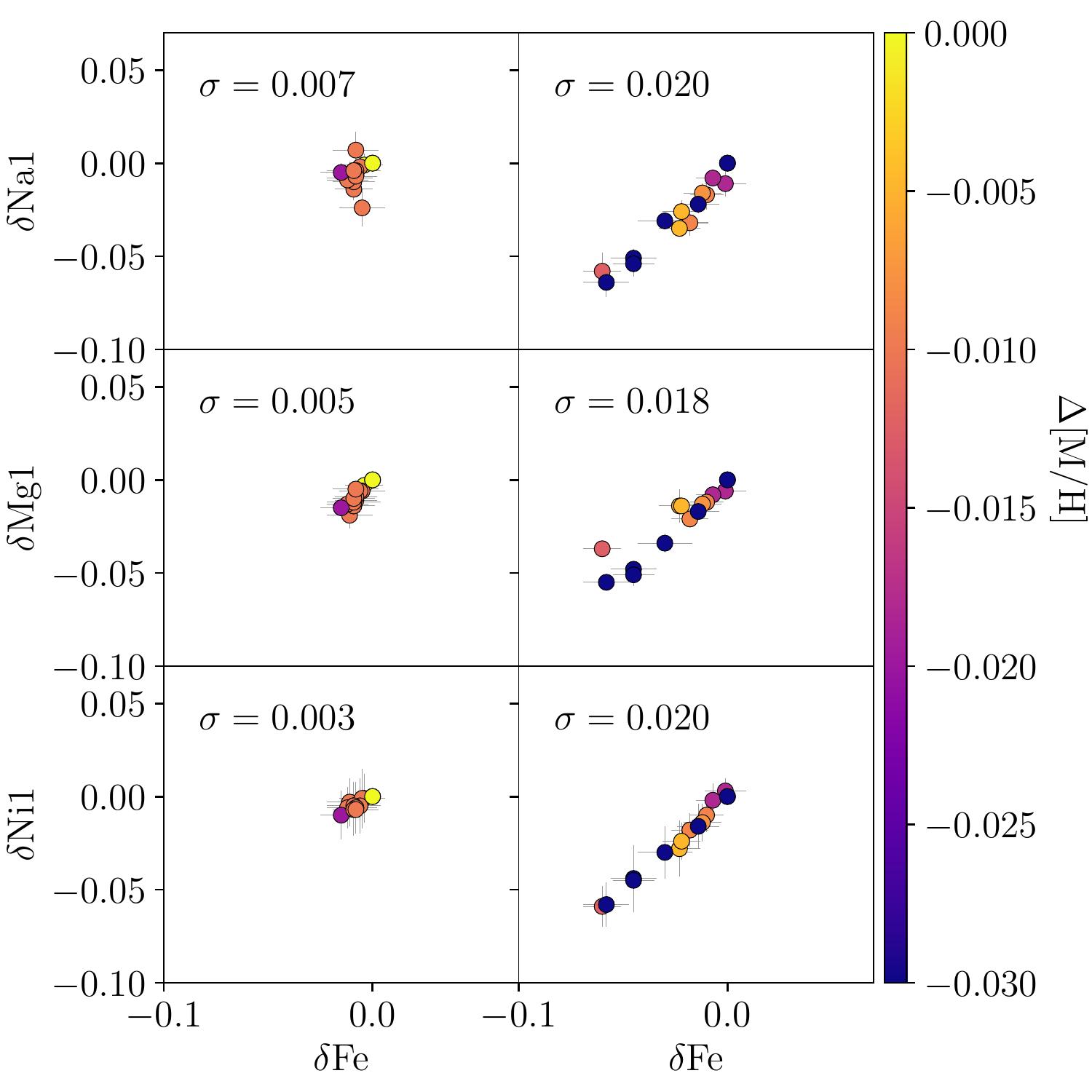}
\caption{Differential abundances of Na, Mg, and Ni as a function of Fe abundance, for the simulated homogeneous (left) and inhomogeneous (right) OCs. The dispersions in the differential abundances are indicated in each panel. The points are colored according to the difference between: recovered and input metallicity.}\label{fig:synth_OC}
\end{figure}

We plot three examples of the resulting differential abundances of the two cases as a function of the Fe differential abundance in Fig.~\ref{fig:synth_OC}. The overall abundance dispersions are one order of magnitude larger in the inhomogeneous case compared to the homogeneous one. Moreover, the dispersions in the homogeneous case are usually of the same order as the abundance uncertainties. Tight abundance correlations appear in the inhomogeneous cluster, and they appear for any given pair of chemical species, although we only show few cases in the figure. For the homogeneous case, a hint of correlation is seen for some elements (e.g. Mg and Ni), but it can be explained by the wrong choice of the metallicity of the model, indicated by the color of the points. This means that a slightly smaller abundance in all elements will be expected if the chosen atmosphere model has a lower (and wrong) overall metallicity. On the contrary, in the inhomogeneous case, there is no indication that the retrieved smaller differential abundances are due to a wrong metallicity in the model.

Compared with the results on Fig~\ref{fig:XHXH_OCs}, the analyzed stars in the three clusters resemble the inhomogeneous case in all chemical species we have been able to perform a fit. We conclude that the analyzed OCs present signs of chemical inhomogeneity at a level of $\sim$0.02-0.03 dex with no clear dependence on the chemical species.

\subsection{Chemical tagging}
\begin{figure}
\centering
\includegraphics[width=0.5\textwidth]{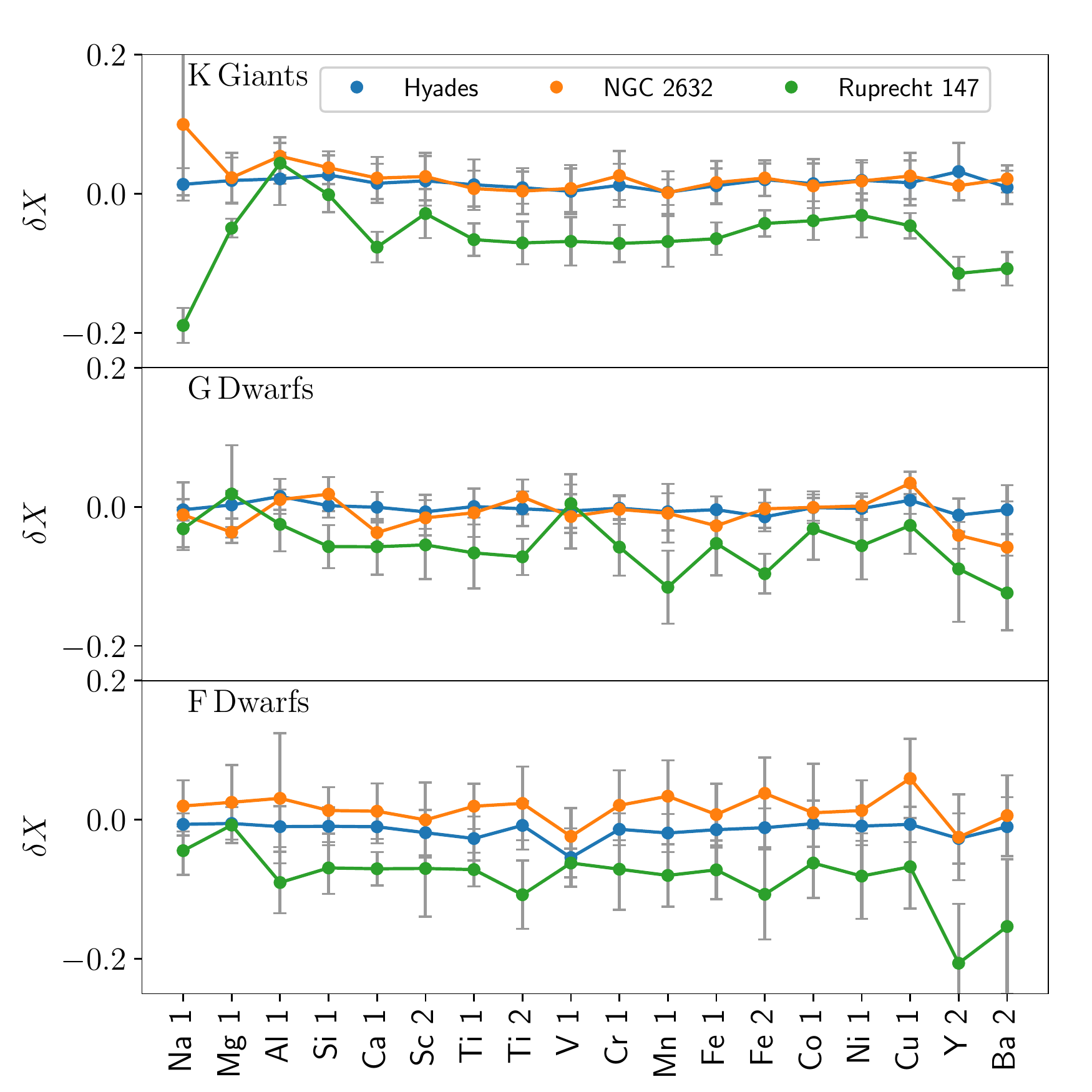}
\caption{Differential chemical abundances (weighted average) of all analyzed elements of: K giants (top), G dwarfs (middle) and F dwarfs (bottom).} \label{fig:abusOC_chemtag}
\end{figure}

Our sample of stars is well suited to investigate the chemical tagging using differential abundance computation. 
We plot in Fig~\ref{fig:abusOC_chemtag} the weighted average abundances and dispersions for three groups of stars: K giants, G dwarfs\footnote{we restrict this group to $5900>\teff>5500$ K since only the Hyades have a representative number of stars cooler than that.} and F dwarfs. We only include elements that we are able to measure in at least one star in the three groups and the three clusters.

G dwarfs and K giants in NGC~2632 and the Hyades have an equivalent chemical signature for almost all elements. In terms of the chemical evolution of the Galaxy, this is consistent with the two OCs having similar ages, as computed in the literature on several occasions using isochrones. However, still some discrepancies are found in the age value, usually quoted between 600-800 Myr depending on methodology and models used \citep[e.g.][]{Brandt+2015,Gossage+2018}.

Instead, the F dwarfs group for NGC~2632 appears enhanced in all elements compared to the Hyades stars, but with large dispersions. This is mainly produced by few stars in this cluster in the range $6000<\teff<6200$ K. They give very consistent abundances among them, but $\sim0.05$ dex larger than the other stars in the F dwarfs group, the latter matching the abundances of the Hyades. We have checked that this is not due to low S/N of the spectra, or to a particular instrument configuration, and also does not depend on the choice of the reference star. This effect can be related to enhancement by convection of stars near the turnoff since at this temperature range is where the mass of the convective zone increases \citep{Pinsonneault+2001}. Another possibility is that this is caused by atomic diffusion, for which some models predict a dip in the abundances in this region of the HR diagram \citep[e.g.][]{Souto+2019}. Further investigation is needed to reach a conclusion.

In the three panels, we can differentiate the chemical signature of Ruprecht~147 in comparison with the other two clusters, where the abundances for Ruprecht~147 are systematically lower by typically $0.05$ to $0.07$ dex. In the case of giants, uncertainties are lower so this allows to have a better distinction of the clusters. This chemical separation is explained with the different age of this cluster, around 2.5 Gyr (see Table~\ref{tab:clusters}). We remark the particular case of lower \ion{Na}{I} for the giants in Ruprecht~147 (compared to the Hyades). We attribute this as an effect of internal mixing in the surface of massive giants described by the stellar evolutionary models of e.g. \citet[][]{Lagarde+2012}. Similar Na differences were identified in \citet{Smiljanic+2016} for red clump stars in clusters depending on stellar mass, where stars more massive than $\sim2\,M_{\odot}$ can present overabundances up to 0.2 dex.
Ruprecht~147 is much older than the Hyades, therefore, its giants are less massive: checking with isochrones they are about $1.6\,M_{\odot}$, compared to $2.5\,M_{\odot}$ in the Hyades. So in our case, we see Na underabundances since we are analyzing differentially with respect to the Hyades. Non-LTE effects could also play a role in our analysis since the giants in Ruprecht~147 have slightly different atmospheric parameters with respect to the Hyades.

\begin{figure}
\centering
\includegraphics[width=0.5\textwidth]{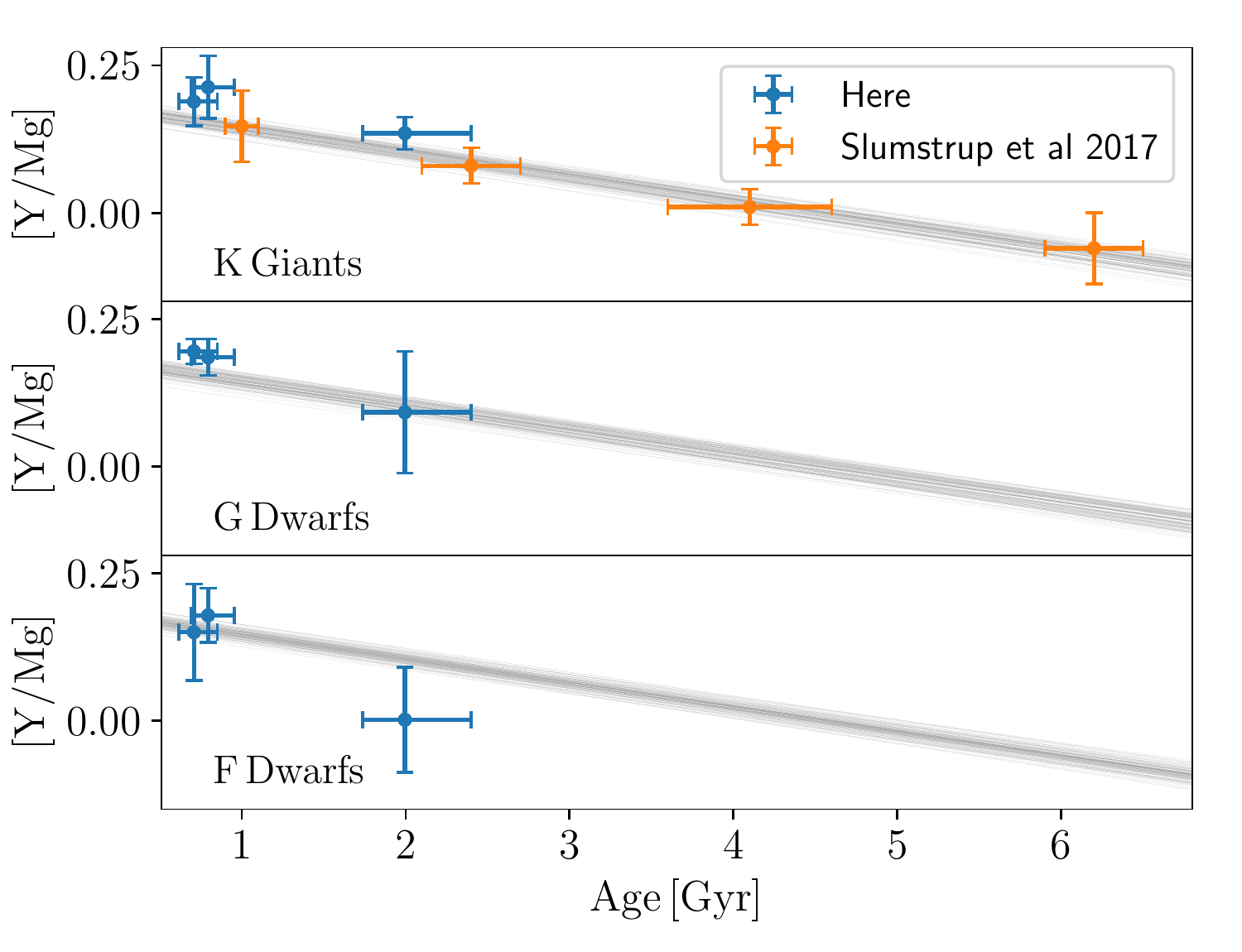}
\caption{[Y/Mg] abundances with respect to the age of the three clusters studied in this work (blue), and the clusters studied in \citet{Slumstrup+2017} (orange). We plot the results of the different spectral types in the three planets. We overplot the relation found by \citet{TucciMaia+2016} in gray.} \label{fig:YMg_age}
\end{figure}

We find peculiarities in several elements, for example in \ion{Y}{II} and \ion{Ba}{II}, which differs in Ruprecht~147 compared to the other two clusters. This is probably due to the different nucleosynthetic origin of these two elements, which are basically produced by neutron captures via the s-process. The impact of the age in the chemical abundances of these types of elements motivates their use to construct "chemical clocks" \citep[e.g.][]{TucciMaia+2016}. In Fig.~\ref{fig:YMg_age} we plot the derived [Y/Mg] as a function of the age reported by \citet{GaiaCollaborationB+2018}. We have computed [Y/Mg] from our differential abundances with respect to the Hyades, and adding the absolute [Y/H] and [Mg/H] abundance of the Hyades with respect to the Sun (Table~\ref{tab:cluster_abundances}). Our uncertainties for Ruprecht 147 dwarfs are quite large for these two elements, mainly because the \gls{SNR} are the lowest for these spectra. For comparison purposes, we overplot the empirical relation derived by \citet{TucciMaia+2016} using solar twins taking into account the uncertainties that they quote. We have added to the plot the four clusters analyzed by \citet{Slumstrup+2017}, where they compute the [Y/Mg] of 1 to 3 red giant stars in each cluster. The points of both studies are compatible with the empirical relation at $1\sigma$. This result is interesting in the context of Galactic archaeology since it is proving that helium-core burning giants follow this relation similarly to dwarfs.

\subsection{Tidal tails}\label{sec:tails}

In our initial sample, we included ten and three stars from the tidal tails of the Hyades and NGC~2632, respectively. Several of them were removed from the sample in the membership refinement in Sect.\ref{sec:galvel}. We have recovered the differential abundances of those that could be analyzed differentially\footnote{Three tidal tail stars rejected as outliers could not be analyzed differentially because they fall out of the group limits} indicated in Fig.~\ref{fig:groups} to compare their chemical signature with respect to the cluster. 

\begin{figure}
\centering
\includegraphics[width=0.5\textwidth]{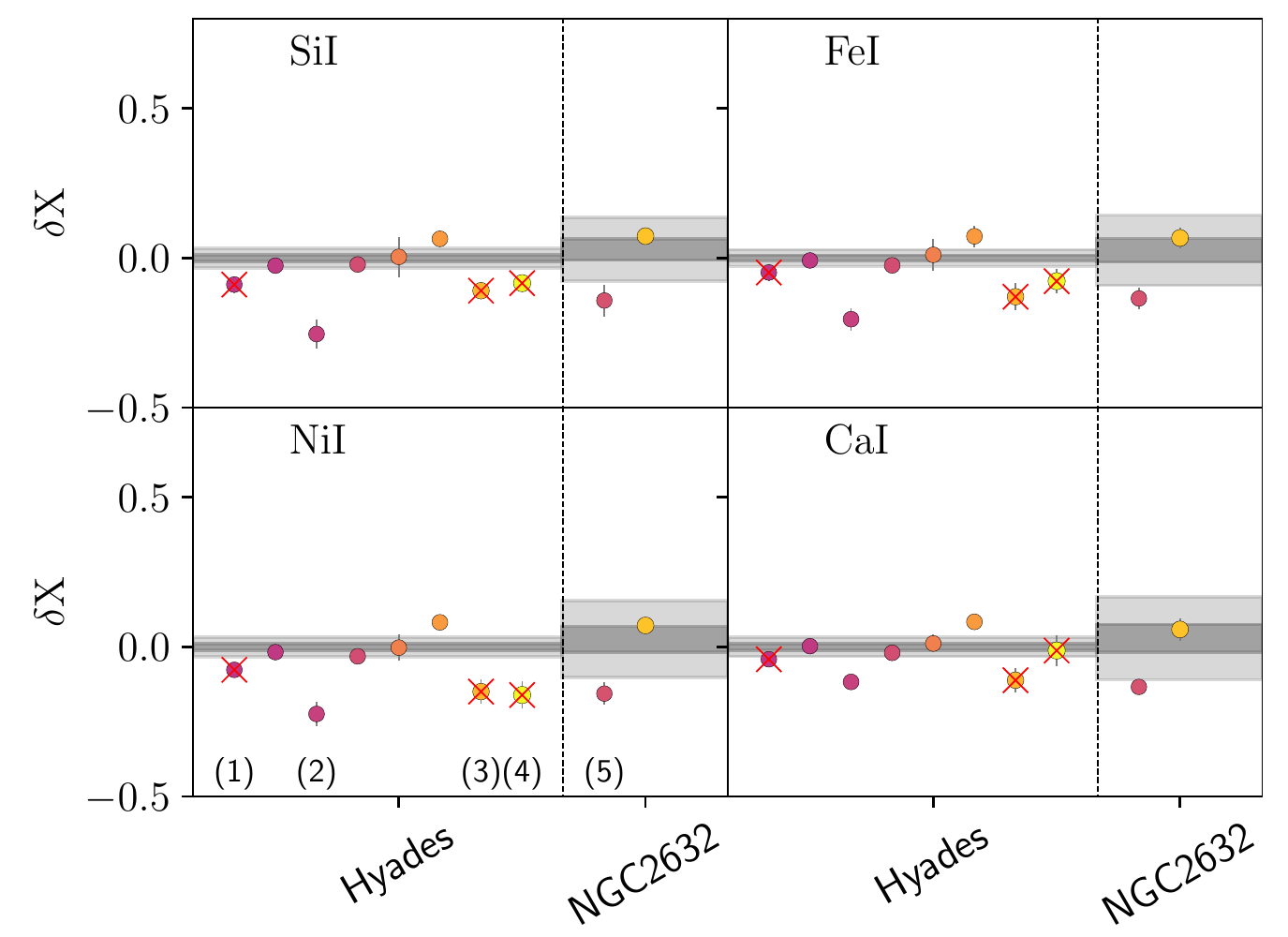}
\caption{Differential chemical abundances of Si, Fe, Ni and Ca of the analyzed stars in the tidal tails of the Hyades and NGC~2632. The color code is the same as in Fig.~\ref{fig:differential_abus}. Red crosses indicate stars identified as kinematic outliers. Shadowed regions represent the cluster signature: $1\sigma$ (dark grey) and $3\sigma$ (light grey) dispersion. Numbered stars are discussed in the text.}\label{fig:abusOC_tails}
\end{figure}

We are able to compute differential abundances for eight stars in the Hyades, and two in NGC~2632.
As an example, we plot in Fig.~\ref{fig:abusOC_tails} abundances of \ion{Fe}{I}, \ion{Si}{I}, \ion{Ni}{I} and \ion{Mg}{I} of the tidal tails stars, highlighting those rejected as outliers with crosses. The numbered stars in the plot are discussed below:

\begin{itemize}
\item The stars 2 and 5, are out of the $3\sigma$ signature of the respective clusters in almost all the chemical species. These were not identified as outliers in kinematics in Sect.\ref{sec:galvel}, but they are probably not members.
\item Star 3 is out of $3\sigma$ in 10 of the chemical species, and in the other cases is out of $1\sigma$. It was identified as an outlier and it is probably a non-member.
\item Star 4 is out of $3\sigma$ in 7 elements and in most of the other elements, it is out of $1\sigma$. It was identified as an outlier and can be a non-member.
\item Star 1 is inside of $3\sigma$ in most of the elements. It was identified as an outlier in Sect.\ref{sec:galvel} but the chemical signature is mainly compatible. It could be that this star is simply in the tail of the kinematic distribution of the cluster and that is why we rejected it as an outlier.
\end{itemize}

According to the results merging kinematics and chemistry, we have discarded as members four (stars 2, 3, 4 and 5) out of 10 analyzed stars. So we conclude that there exists significant contamination when selecting member stars in the outskirts of the clusters using only kinematics. This can be disentangled when looking at the detailed chemistry using differential analysis.

\section{Conclusions}
In this work, we derived radial velocities, atmospheric parameters, absolute and differential chemical abundances of 22 chemical species for stars in three nearby \glspl{OC}: the Hyades, NGC~2632 (Praesepe) and Ruprecht~147. We investigated the possibilities of differential analysis to analyze the homogeneity of nearby clusters and their tidal tails, and the chemical tagging using stars in different evolutionary states.
We used the most recent membership of the three clusters which are very accurate thanks to \emph{Gaia} DR2 exquisite astrometry at small distances ($d<300$ pc). We queried the public archives to look for high resolution and high \gls{SNR} spectra for the targeted stars, and we also performed our own observational programs. We were able to analyze spectra of 62, 22 and 24 stars in the Hyades, NGC~2632 and Ruprecht~147, respectively.

To ensure a sample of stars without contaminants, we did a membership refinement computing total Galactic velocities from the \emph{Gaia} DR2 proper motions and parallaxes, and the obtained radial velocities from our pipeline. We rejected several outliers previously classified as members, which can be contaminants, spectroscopic binaries, or stars in the tails of the Galactic velocity distribution. We obtained atmospheric parameters, and chemical abundances with respect to the Sun ([X/H]). We show that bracket abundances depend on the atmospheric parameters for certain elements (e.g. Na, Mg, V, Mg, and La). We attribute this to a mix of non-LTE effects, physical changes of the abundances in the stellar atmosphere depending on evolutionary stage, and systematics in the analysis.

Strictly line-by-line differential abundances were computed for a sample of 92 stars in eight groups of twin stars, using as reference a star from the Hyades. The precision of the derived differential abundances was between 0.01-0.02 dex, with the exception of the heavy elements. A great effort was done to quantify the uncertainties of the analysis, internally and externally, following the recommendations of \citet{Jofre+2019}. We compared the results of different spectra of the same star analyzed independently, we tested the effects of the uncertainties in the atmospheric parameters, and we did an extensive comparison of the stellar parameters and abundances retrieved for the \gls{GBS}.

Differential abundances did not present a dependence with spectral type. An exception to this is Ruprecht~147 giants which showed underabundances of Na of almost 0.2 dex with respect to the Hyades. Several studies propose that red giants more massive than $2M_{\odot}$ can present Na overabundances due to internal mixing \citep[e.g.][]{Lagarde+2012}. Because of the age difference of the Hyades and Ruprecht 147 their giants have $\sim2.5M_{\odot}$ and $\sim1.6M_{\odot}$, and so this is consistent with the found difference.

We investigated the level of chemical homogeneity of the three clusters using the set of differential abundances obtained after combining different kinds of stars and instruments. We obtain large amplitudes in all chemical species, compared with our uncertainties, and dispersions of the order 0.02-0.03 dex in the Hyades. Moreover, very significant correlations are found for almost all pairs of elements with low dispersion. We used our pipeline to analyze a homogeneous vs inhomogeneous synthetic cluster, showing that correlations appear when stars have some level of chemical inhomogeneity. This confirms the analysis done previously by \citet{Liu+2016} for the Hyades, with three times more stars. Most of the slopes in the correlations between abundance pairs are found compatible and around 1, showing that this is due to an overall zero-point difference in the stars. For NGC~2632 and Ruprecht~147 we deal with larger uncertainties, especially for the dwarfs, which make the dispersions and amplitudes in abundances to be larger than those of the Hyades. We obtain also signs of abundance correlations in several abundance pairs, which suggests some chemical inhomogeneity also.
s

Chemical tagging was analyzed for three spectral types of stars: F dwarfs, G dwarfs, and K giants. A clear difference in the chemical signature is seen for Ruprecht~147 with respect to the other two clusters, due to the different age ($\sim$2.5 Gyr vs $\sim$700 Myr of the Hyades). The largest difference is seen for the giants, where we benefit from lower uncertainties in the abundances. A particularly large difference in the signatures is found for the s-process elements Y and Ba. The Hyades and NGC~2632 are indistinguishable in almost all elements, which is consistent with the two clusters having the same age. We remark the enhancement of about $\sim0.05$ dex found for a few F dwarfs in the temperature range $6000<\teff<6200$ K. The effect needs further investigation, it can be related to the change of the convective zones, and/or to atomic diffusion.

Finally, we analyzed in detail the differential chemical abundances of the stars identified in the tidal tails of the Hyades (8 stars) and NGC~2632 (2 stars), with respect to the cluster chemical signature. Four out of the ten stars do not seem members according to the chemistry. Two of them can be identified as outliers when analyzing total Galactic velocities using radial velocity coming from high-resolution spectroscopy. We conclude that a lot of contamination exists when selecting members in the outskirts of a cluster, based only on kinematics.

\begin{acknowledgements}
We gratefully thank the anonymous referee for providing comments that have improved the quality of the work.

This work has made use of data from the European Space Agency (ESA) mission \emph{Gaia} (\url{http://www.cosmos.esa.int/gaia}), processed by the \emph{Gaia} Data Processing and Analysis Consortium (DPAC, \url{http://www.cosmos.esa.int/web/gaia/dpac/consortium}). We acknowledge the \emph{Gaia} Project Scientist Support Team and the \emph{Gaia} DPAC. Funding for the DPAC has been provided by national institutions, in particular, the institutions participating in the \emph{Gaia} Multilateral Agreement.
This research made extensive use of the SIMBAD database, and the VizieR catalog access tool operated at the CDS, Strasbourg, France, and of NASA Astrophysics Data System Bibliographic Services.
This research has made use of Astropy \citep{Astropy2013}, Topcat \citep{Taylor2005}.

L.C., C.S., and Y.T. acknowledge support from "programme national de physique stellaire" (PNPS) and from the "programme national cosmologie et galaxies" (PNCG) of CNRS/INSU. U.H. acknowledges support from the Swedish National Space Agency (SNSA/Rymdstyrelsen).

\end{acknowledgements}

\bibliographystyle{aa} 
\bibliography{biblio_v4,biblio2_v1}

\begin{thebibliography}{68}
\expandafter\ifx\csname natexlab\endcsname\relax\def\natexlab#1{#1}\fi

\bibitem[{{Astropy Collaboration} {et~al.}(2013){Astropy Collaboration},
  {Robitaille}, {Tollerud}, {Greenfield}, {Droettboom}, {Bray}, {Aldcroft},
  {Davis}, {Ginsburg}, {Price-Whelan}, {Kerzendorf}, {Conley}, {Crighton},
  {Barbary}, {Muna}, {Ferguson}, {Grollier}, {Parikh}, {Nair}, {Unther},
  {Deil}, {Woillez}, {Conseil}, {Kramer}, {Turner}, {Singer}, {Fox}, {Weaver},
  {Zabalza}, {Edwards}, {Azalee Bostroem}, {Burke}, {Casey}, {Crawford},
  {Dencheva}, {Ely}, {Jenness}, {Labrie}, {Lim}, {Pierfederici}, {Pontzen},
  {Ptak}, {Refsdal}, {Servillat}, \& {Streicher}}]{Astropy2013}
{Astropy Collaboration}, {Robitaille}, T.~P., {Tollerud}, E.~J., {et~al.} 2013,
  \aap, 558, A33

\bibitem[{{Bergemann} {et~al.}(2019){Bergemann}, {Gallagher}, {Eitner},
  {Bautista}, {Collet}, {Yakovleva}, {Mayriedl}, {Plez}, {Carlsson},
  {Leenaarts}, {Belyaev}, \& {Hansen}}]{Bergemann+2019}
{Bergemann}, M., {Gallagher}, A.~J., {Eitner}, P., {et~al.} 2019, arXiv
  e-prints, arXiv:1905.05200

\bibitem[{{Blanco-Cuaresma}(2019)}]{BlancoCuaresma2019}
{Blanco-Cuaresma}, S. 2019, \mnras, 486, 2075

\bibitem[{{Blanco-Cuaresma} \& {Fraix-Burnet}(2018)}]{BlancoCuaresma+2018}
{Blanco-Cuaresma}, S. \& {Fraix-Burnet}, D. 2018, \aap, 618, A65

\bibitem[{{Blanco-Cuaresma} {et~al.}(2014{\natexlab{a}}){Blanco-Cuaresma},
  {Soubiran}, {Heiter}, \& {Jofr{\'e}}}]{BlancoCuaresma+2014}
{Blanco-Cuaresma}, S., {Soubiran}, C., {Heiter}, U., \& {Jofr{\'e}}, P.
  2014{\natexlab{a}}, A\&A, 569, A111

\bibitem[{{Blanco-Cuaresma} {et~al.}(2014{\natexlab{b}}){Blanco-Cuaresma},
  {Soubiran}, {Jofr{\'e}}, \& {Heiter}}]{Blanco+2014}
{Blanco-Cuaresma}, S., {Soubiran}, C., {Jofr{\'e}}, P., \& {Heiter}, U.
  2014{\natexlab{b}}, A\&A, 566, A98

\bibitem[{{Boesgaard} {et~al.}(2013){Boesgaard}, {Roper}, \&
  {Lum}}]{Boesgaard+2013}
{Boesgaard}, A.~M., {Roper}, B.~W., \& {Lum}, M.~G. 2013, \apj, 775, 58

\bibitem[{{Bossini} {et~al.}(2019){Bossini}, {Vallenari}, {Bragaglia},
  {Cantat-Gaudin}, {Sordo}, {Balaguer-N{\'u}{\~n}ez}, {Jordi}, {Moitinho},
  {Soubiran}, {Casamiquela}, {Carrera}, \& {Heiter}}]{Bossini+2019}
{Bossini}, D., {Vallenari}, A., {Bragaglia}, A., {et~al.} 2019, \aap, 623, A108

\bibitem[{{Bragaglia} {et~al.}(2018){Bragaglia}, {Fu}, {Mucciarelli},
  {Andreuzzi}, \& {Donati}}]{Bragaglia+2018}
{Bragaglia}, A., {Fu}, X., {Mucciarelli}, A., {Andreuzzi}, G., \& {Donati}, P.
  2018, \aap, 619, A176

\bibitem[{{Brandt} \& {Huang}(2015)}]{Brandt+2015}
{Brandt}, T.~D. \& {Huang}, C.~X. 2015, \apj, 807, 24

\bibitem[{{Bressan} {et~al.}(2012){Bressan}, {Marigo}, {Girardi}, {Salasnich},
  {Dal Cero}, {Rubele}, \& {Nanni}}]{Bressan+2012}
{Bressan}, A., {Marigo}, P., {Girardi}, L., {et~al.} 2012, \mnras, 427, 127

\bibitem[{{Cantat-Gaudin} {et~al.}(2018){Cantat-Gaudin}, {Jordi}, {Vallenari},
  {Bragaglia}, {Balaguer-N{\'u}{\~n}ez}, {Soubiran}, {Bossini}, {Moitinho},
  {Castro-Ginard}, {Krone-Martins}, {Casamiquela}, {Sordo}, \&
  {Carrera}}]{Cantat-gaudin+2018}
{Cantat-Gaudin}, T., {Jordi}, C., {Vallenari}, A., {et~al.} 2018, \aap, 618,
  A93

\bibitem[{{Carlberg}(2014)}]{Carlberg2014}
{Carlberg}, J.~K. 2014, \aj, 147, 138

\bibitem[{{Casamiquela} {et~al.}(2019){Casamiquela}, {Blanco-Cuaresma},
  {Carrera}, {Balaguer-N{\'u}{\~n}ez}, {Jordi}, {Anders}, {Chiappini},
  {Carbajo-Hijarrubia}, {Aguado}, {del Pino}, {D{\'\i}az-P{\'e}rez}, {Gallart},
  \& {Pancino}}]{Casamiquela+2019}
{Casamiquela}, L., {Blanco-Cuaresma}, S., {Carrera}, R., {et~al.} 2019, arXiv
  e-prints, arXiv:1909.05865

\bibitem[{{Castelli} \& {Kurucz}(2004)}]{Castelli+2004}
{Castelli}, F. \& {Kurucz}, R.~L. 2004, ArXiv Astrophysics e-prints
  [\eprint{astro-ph/0405087}]

\bibitem[{{Curtis} {et~al.}(2013){Curtis}, {Wolfgang}, {Wright}, {Brewer}, \&
  {Johnson}}]{Curtis+2013}
{Curtis}, J.~L., {Wolfgang}, A., {Wright}, J.~T., {Brewer}, J.~M., \&
  {Johnson}, J.~A. 2013, \aj, 145, 134

\bibitem[{{Dias} {et~al.}(2002){Dias}, {Alessi}, {Moitinho}, \&
  {L{\'e}pine}}]{Dias+2002}
{Dias}, W.~S., {Alessi}, B.~S., {Moitinho}, A., \& {L{\'e}pine}, J.~R.~D. 2002,
  A\&A, 389, 871

\bibitem[{{Dotter} {et~al.}(2017){Dotter}, {Conroy}, {Cargile}, \&
  {Asplund}}]{Dotter+2017}
{Dotter}, A., {Conroy}, C., {Cargile}, P., \& {Asplund}, M. 2017, \apj, 840, 99

\bibitem[{{Friel} {et~al.}(2002){Friel}, {Janes}, {Tavarez}, {Scott},
  {Katsanis}, {Lotz}, {Hong}, \& {Miller}}]{Friel+2002}
{Friel}, E.~D., {Janes}, K.~A., {Tavarez}, M., {et~al.} 2002, \aj, 124, 2693

\bibitem[{{Gaia Collaboration} {et~al.}(2018{\natexlab{a}}){Gaia
  Collaboration}, {Babusiaux}, {van Leeuwen}, {Barstow}, {Jordi}, {Vallenari},
  {Bossini}, {Bressan}, {Cantat-Gaudin}, {van Leeuwen}, \&
  et~al.}]{GaiaCollaborationB+2018}
{Gaia Collaboration}, {Babusiaux}, C., {van Leeuwen}, F., {et~al.}
  2018{\natexlab{a}}, \aap, 616, A10

\bibitem[{{Gaia Collaboration} {et~al.}(2018{\natexlab{b}}){Gaia
  Collaboration}, {Brown}, {Vallenari}, {Prusti}, {de Bruijne}, {Babusiaux},
  {Bailer-Jones}, {Biermann}, {Evans}, {Eyer}, \&
  et~al.}]{GaiaCollaboration+2018}
{Gaia Collaboration}, {Brown}, A.~G.~A., {Vallenari}, A., {et~al.}
  2018{\natexlab{b}}, \aap, 616, A1

\bibitem[{{Gebran} {et~al.}(2010){Gebran}, {Vick}, {Monier}, \&
  {Fossati}}]{Gebran+2010}
{Gebran}, M., {Vick}, M., {Monier}, R., \& {Fossati}, L. 2010, \aap, 523, A71

\bibitem[{{Gilmore} {et~al.}(2012){Gilmore}, {Randich}, {Asplund}, {Binney},
  {Bonifacio}, {Drew}, {Feltzing}, {Ferguson}, {Jeffries}, {Micela}, \&
  et~al.}]{Gilmore+2012}
{Gilmore}, G., {Randich}, S., {Asplund}, M., {et~al.} 2012, The Messenger, 147,
  25

\bibitem[{{Gossage} {et~al.}(2018){Gossage}, {Conroy}, {Dotter}, {Choi},
  {Rosenfield}, {Cargile}, \& {Dolphin}}]{Gossage+2018}
{Gossage}, S., {Conroy}, C., {Dotter}, A., {et~al.} 2018, \apj, 863, 67

\bibitem[{{Gray} \& {Corbally}(1994)}]{Gray+1994}
{Gray}, R.~O. \& {Corbally}, C.~J. 1994, \aj, 107, 742

\bibitem[{{Grevesse} {et~al.}(2007){Grevesse}, {Asplund}, \&
  {Sauval}}]{Grevesse+2007}
{Grevesse}, N., {Asplund}, M., \& {Sauval}, A.~J. 2007, \ssr, 130, 105

\bibitem[{{Gustafsson} {et~al.}(2008){Gustafsson}, {Edvardsson}, {Eriksson},
  {J{\o}rgensen}, {Nordlund}, \& {Plez}}]{Gustafsson2008}
{Gustafsson}, B., {Edvardsson}, B., {Eriksson}, K., {et~al.} 2008, A\&A, 486,
  951

\bibitem[{{Hawkins} {et~al.}(2016){Hawkins}, {Masseron}, {Jofr{\'e}},
  {Gilmore}, {Elsworth}, \& {Hekker}}]{Hawkins+2016b}
{Hawkins}, K., {Masseron}, T., {Jofr{\'e}}, P., {et~al.} 2016, \aap, 594, A43

\bibitem[{{Heiter} {et~al.}(2015{\natexlab{a}}){Heiter}, {Jofr{\'e}},
  {Gustafsson}, {Korn}, {Soubiran}, \& {Th{\'e}venin}}]{Heiter+2015}
{Heiter}, U., {Jofr{\'e}}, P., {Gustafsson}, B., {et~al.} 2015{\natexlab{a}},
  \aap, 582, A49

\bibitem[{{Heiter} {et~al.}(2015{\natexlab{b}}){Heiter}, {Lind}, {Asplund},
  {Barklem}, {Bergemann}, {Magrini}, {Masseron}, {Mikolaitis}, {Pickering}, \&
  {Ruffoni}}]{Heiter+2015b}
{Heiter}, U., {Lind}, K., {Asplund}, M., {et~al.} 2015{\natexlab{b}}, \physscr,
  90, 054010

\bibitem[{{Heiter} {et~al.}(2019){Heiter}, {Lind}, {Bergemann}, {Asplund},
  {Mikolaitis}, {Barklem}, {Masseron}, {de Laverny}, {Magrini}, {Edvardsson},
  {J{\"o}nsson}, {Pickering}, {Ryde}, {Bayo}, {Bensby}, {Casey}, {Feltzing},
  {Jofr{\'e}}, {Korn}, {Pancino}, {Damiani}, {Lanzafame}, {Lardo}, {Monaco},
  {Morbidelli}, {Smiljanic}, {Worley}, {Zaggia}, {Randich}, \&
  F.}]{Heiter+2019}
{Heiter}, U., {Lind}, K., {Bergemann}, M., {et~al.} 2019, \aap, submitted

\bibitem[{{Heiter} \& {Luck}(2003)}]{Heiter+2003}
{Heiter}, U. \& {Luck}, R.~E. 2003, \aj, 126, 2015

\bibitem[{{Hogg} {et~al.}(2010){Hogg}, {Bovy}, \& {Lang}}]{Hogg+2010}
{Hogg}, D.~W., {Bovy}, J., \& {Lang}, D. 2010, arXiv e-prints
  [\eprint[arXiv]{1008.4686}]

\bibitem[{{Jofr{\'e}} {et~al.}(2019){Jofr{\'e}}, {Heiter}, \&
  {Soubiran}}]{Jofre+2019}
{Jofr{\'e}}, P., {Heiter}, U., \& {Soubiran}, C. 2019, \araa, 57, 571

\bibitem[{{Jofr{\'e}} {et~al.}(2015){Jofr{\'e}}, {Heiter}, {Soubiran},
  {Blanco-Cuaresma}, {Masseron}, {Nordlander}, {Chemin}, {Worley}, {Van Eck},
  {Hourihane}, {Gilmore}, {Adibekyan}, {Bergemann}, {Cantat-Gaudin},
  {Delgado-Mena}, {Gonz{\'a}lez Hern{\'a}ndez}, {Guiglion}, {Lardo}, {de
  Laverny}, {Lind}, {Magrini}, {Mikolaitis}, {Montes}, {Pancino},
  {Recio-Blanco}, {Sordo}, {Sousa}, {Tabernero}, \& {Vallenari}}]{Jofre+2015}
{Jofr{\'e}}, P., {Heiter}, U., {Soubiran}, C., {et~al.} 2015, \aap, 582, A81

\bibitem[{{Jofr{\'e}} {et~al.}(2014){Jofr{\'e}}, {Heiter}, {Soubiran}, \&
  et~al.}]{Jofre+2014}
{Jofr{\'e}}, P., {Heiter}, U., {Soubiran}, C., \& et~al. 2014, \aap, 564, A133

\bibitem[{{Jofr{\'e}} {et~al.}(2018){Jofr{\'e}}, {Heiter}, {Tucci Maia},
  {Soubiran}, {Worley}, {Hawkins}, {Blanco-Cuaresma}, \&
  {Rodrigo}}]{Jofre+2018}
{Jofr{\'e}}, P., {Heiter}, U., {Tucci Maia}, M., {et~al.} 2018, Research Notes
  of the American Astronomical Society, 2, 152

\bibitem[{{Kharchenko} {et~al.}(2005){Kharchenko}, {Piskunov}, {R{\"o}ser},
  {Schilbach}, \& {Scholz}}]{Karchenko+2005}
{Kharchenko}, N.~V., {Piskunov}, A.~E., {R{\"o}ser}, S., {Schilbach}, E., \&
  {Scholz}, R.-D. 2005, \aap, 438, 1163

\bibitem[{{Kurucz}(1993)}]{Kurucz1993}
{Kurucz}, R. 1993, SYNTHE Spectrum Synthesis Programs and Line Data.~Kurucz
  CD-ROM No.~18.~Cambridge, Mass.: Smithsonian Astrophysical Observatory,
  1993., 18

\bibitem[{{Lagarde} {et~al.}(2012){Lagarde}, {Decressin}, {Charbonnel},
  {Eggenberger}, {Ekstr{\"o}m}, \& {Palacios}}]{Lagarde+2012}
{Lagarde}, N., {Decressin}, T., {Charbonnel}, C., {et~al.} 2012, \aap, 543,
  A108

\bibitem[{{Lind} {et~al.}(2011){Lind}, {Asplund}, {Barklem}, \&
  {Belyaev}}]{Lind+2011}
{Lind}, K., {Asplund}, M., {Barklem}, P.~S., \& {Belyaev}, A.~K. 2011, \aap,
  528, A103

\bibitem[{{Liu} {et~al.}(2019){Liu}, {Asplund}, {Yong}, {Feltzing}, {Dotter},
  {Mel{\'e}ndez}, \& {Ram{\'\i}rez}}]{Liu+2019}
{Liu}, F., {Asplund}, M., {Yong}, D., {et~al.} 2019, \aap, 627, A117

\bibitem[{{Liu} {et~al.}(2016{\natexlab{a}}){Liu}, {Asplund}, {Yong},
  {Mel{\'e}ndez}, {Ram{\'\i}rez}, {Karakas}, {Carlos}, \& {Marino}}]{Liu+2016b}
{Liu}, F., {Asplund}, M., {Yong}, D., {et~al.} 2016{\natexlab{a}}, \mnras, 463,
  696

\bibitem[{{Liu} {et~al.}(2016{\natexlab{b}}){Liu}, {Yong}, {Asplund},
  {Ram{\'{\i}}rez}, \& {Mel{\'e}ndez}}]{Liu+2016}
{Liu}, F., {Yong}, D., {Asplund}, M., {Ram{\'{\i}}rez}, I., \& {Mel{\'e}ndez},
  J. 2016{\natexlab{b}}, \mnras, 457, 3934

\bibitem[{{Mahdi} {et~al.}(2016){Mahdi}, {Soubiran}, {Blanco-Cuaresma}, \&
  {Chemin}}]{Mahdi+2016}
{Mahdi}, D., {Soubiran}, C., {Blanco-Cuaresma}, S., \& {Chemin}, L. 2016, \aap,
  587, A131

\bibitem[{{Majewski} {et~al.}(2017){Majewski}, {Schiavon}, {Frinchaboy},
  {Allende Prieto}, {Barkhouser}, {Bizyaev}, {Blank}, {Brunner}, {Burton},
  {Carrera}, {Chojnowski}, {Cunha}, {Epstein}, {Fitzgerald}, {Garc{\'{\i}}a
  P{\'e}rez}, {Hearty}, {Henderson}, {Holtzman}, {Johnson}, {Lam}, {Lawler},
  {Maseman}, {M{\'e}sz{\'a}ros}, {Nelson}, {Nguyen}, {Nidever}, {Pinsonneault},
  {Shetrone}, {Smee}, {Smith}, {Stolberg}, {Skrutskie}, {Walker}, {Wilson},
  {Zasowski}, {Anders}, {Basu}, {Beland}, {Blanton}, {Bovy}, {Brownstein},
  {Carlberg}, {Chaplin}, {Chiappini}, {Eisenstein}, {Elsworth}, {Feuillet},
  {Fleming}, {Galbraith-Frew}, {Garc{\'{\i}}a}, {Garc{\'{\i}}a-Hern{\'a}ndez},
  {Gillespie}, {Girardi}, {Gunn}, {Hasselquist}, {Hayden}, {Hekker}, {Ivans},
  {Kinemuchi}, {Klaene}, {Mahadevan}, {Mathur}, {Mosser}, {Muna}, {Munn},
  {Nichol}, {O'Connell}, {Parejko}, {Robin}, {Rocha-Pinto}, {Schultheis},
  {Serenelli}, {Shane}, {Silva Aguirre}, {Sobeck}, {Thompson}, {Troup},
  {Weinberg}, \& {Zamora}}]{Majewski+2017}
{Majewski}, S.~R., {Schiavon}, R.~P., {Frinchaboy}, P.~M., {et~al.} 2017, \aj,
  154, 94

\bibitem[{{Marigo} {et~al.}(2017){Marigo}, {Girardi}, {Bressan}, {Rosenfield},
  {Aringer}, {Chen}, {Dussin}, {Nanni}, {Pastorelli}, {Rodrigues}, {Trabucchi},
  {Bladh}, {Dalcanton}, {Groenewegen}, {Montalb{\'a}n}, \&
  {Wood}}]{Marigo+2017}
{Marigo}, P., {Girardi}, L., {Bressan}, A., {et~al.} 2017, \apj, 835, 77

\bibitem[{{Meingast} \& {Alves}(2019)}]{Meingast+2019}
{Meingast}, S. \& {Alves}, J. 2019, Astronomy and Astrophysics, 621, L3

\bibitem[{{Mel{\'e}ndez} {et~al.}(2009){Mel{\'e}ndez}, {Asplund}, {Gustafsson},
  \& {Yong}}]{Melendez+2009b}
{Mel{\'e}ndez}, J., {Asplund}, M., {Gustafsson}, B., \& {Yong}, D. 2009, \apjl,
  704, L66

\bibitem[{{Nielsen} {et~al.}(2013){Nielsen}, {Gizon}, {Schunker}, \&
  {Karoff}}]{Nielsen+2013}
{Nielsen}, M.~B., {Gizon}, L., {Schunker}, H., \& {Karoff}, C. 2013, \aap, 557,
  L10

\bibitem[{{Nissen}(2015)}]{Nissen2015}
{Nissen}, P.~E. 2015, \aap, 579, A52

\bibitem[{{{\"O}nehag} {et~al.}(2014){{\"O}nehag}, {Gustafsson}, \&
  {Korn}}]{Onehag+2014}
{{\"O}nehag}, A., {Gustafsson}, B., \& {Korn}, A. 2014, \aap, 562, A102

\bibitem[{{Pinsonneault} {et~al.}(2001){Pinsonneault}, {DePoy}, \&
  {Coffee}}]{Pinsonneault+2001}
{Pinsonneault}, M.~H., {DePoy}, D.~L., \& {Coffee}, M. 2001, \apjl, 556, L59

\bibitem[{{Randich} {et~al.}(2013){Randich}, {Gilmore}, \& {Gaia-ESO
  Consortium}}]{Randich+2013}
{Randich}, S., {Gilmore}, G., \& {Gaia-ESO Consortium}. 2013, The Messenger,
  154, 47

\bibitem[{{Reggiani} {et~al.}(2017){Reggiani}, {Mel{\'e}ndez}, {Kobayashi},
  {Karakas}, \& {Placco}}]{Reggiani+2017}
{Reggiani}, H., {Mel{\'e}ndez}, J., {Kobayashi}, C., {Karakas}, A., \&
  {Placco}, V. 2017, \aap, 608, A46

\bibitem[{{Riedel} {et~al.}(2017){Riedel}, {Blunt}, {Lambrides}, {Rice},
  {Cruz}, \& {Faherty}}]{Riedel+2017}
{Riedel}, A.~R., {Blunt}, S.~C., {Lambrides}, E.~L., {et~al.} 2017, \aj, 153,
  95

\bibitem[{{R{\"o}ser} \& {Schilbach}(2019)}]{Roser+2019b}
{R{\"o}ser}, S. \& {Schilbach}, E. 2019, Astronomy and Astrophysics, 627, A4

\bibitem[{{R{\"o}ser} {et~al.}(2019){R{\"o}ser}, {Schilbach}, \&
  {Goldman}}]{Roser+2019}
{R{\"o}ser}, S., {Schilbach}, E., \& {Goldman}, B. 2019, \aap, 621, L2

\bibitem[{{Slumstrup} {et~al.}(2017){Slumstrup}, {Grundahl}, {Brogaard},
  {Thygesen}, {Nissen}, {Jessen-Hansen}, {Van Eylen}, \&
  {Pedersen}}]{Slumstrup+2017}
{Slumstrup}, D., {Grundahl}, F., {Brogaard}, K., {et~al.} 2017, \aap, 604, L8

\bibitem[{{Smiljanic} {et~al.}(2016){Smiljanic}, {Romano}, {Bragaglia},
  {Donati}, {Magrini}, {Friel}, {Jacobson}, {Randich}, {Ventura}, \&
  {Lind}}]{Smiljanic+2016}
{Smiljanic}, R., {Romano}, D., {Bragaglia}, A., {et~al.} 2016, \aap, 589, A115

\bibitem[{{Soubiran} {et~al.}(2018){Soubiran}, {Cantat-Gaudin},
  {Romero-G{\'o}mez}, {Casamiquela}, {Jordi}, {Vallenari}, {Antoja},
  {Balaguer-N{\'u}{\~n}ez}, {Bossini}, {Bragaglia}, {Carrera}, {Castro-Ginard},
  {Figueras}, {Heiter}, {Katz}, {Krone-Martins}, {Le Campion}, {Moitinho}, \&
  {Sordo}}]{Soubiran+2018}
{Soubiran}, C., {Cantat-Gaudin}, T., {Romero-G{\'o}mez}, M., {et~al.} 2018,
  \aap, 619, A155

\bibitem[{{Souto} {et~al.}(2019){Souto}, {Allende Prieto}, {Cunha},
  {Pinsonneault}, {Smith}, {Garcia-Dias}, {Bovy}, {Garc{\'\i}a-Hern{\'a}ndez},
  {Holtzman}, {Johnson}, {J{\"o}nsson}, {Majewski}, {Shetrone}, {Sobeck},
  {Zamora}, {Pan}, \& {Nitschelm}}]{Souto+2019}
{Souto}, D., {Allende Prieto}, C., {Cunha}, K., {et~al.} 2019, \apj, 874, 97

\bibitem[{{Souto} {et~al.}(2018){Souto}, {Cunha}, {Smith}, {Allende Prieto},
  {Garc{\'{\i}}a-Hern{\'a}ndez}, {Pinsonneault}, {Holzer}, {Frinchaboy},
  {Holtzman}, {Johnson}, {J{\"o}nsson}, {Majewski}, {Shetrone}, {Sobeck},
  {Stringfellow}, {Teske}, {Zamora}, {Zasowski}, {Carrera}, {Stassun},
  {Fernandez-Trincado}, {Villanova}, {Minniti}, \& {Santana}}]{Souto+2018}
{Souto}, D., {Cunha}, K., {Smith}, V.~V., {et~al.} 2018, \apj, 857, 14

\bibitem[{{Spina} {et~al.}(2018){Spina}, {Mel{\'e}ndez}, {Casey}, {Karakas}, \&
  {Tucci-Maia}}]{Spina+2018}
{Spina}, L., {Mel{\'e}ndez}, J., {Casey}, A.~R., {Karakas}, A.~I., \&
  {Tucci-Maia}, M. 2018, \apj, 863, 179

\bibitem[{{Taylor}(2005)}]{Taylor2005}
{Taylor}, M.~B. 2005, in Astronomical Society of the Pacific Conference Series,
  Vol. 347, Astronomical Data Analysis Software and Systems XIV, ed.
  P.~{Shopbell}, M.~{Britton}, \& R.~{Ebert}, 29

\bibitem[{{Tucci Maia} {et~al.}(2016){Tucci Maia}, {Ram{\'{\i}}rez},
  {Mel{\'e}ndez}, {Bedell}, {Bean}, \& {Asplund}}]{TucciMaia+2016}
{Tucci Maia}, M., {Ram{\'{\i}}rez}, I., {Mel{\'e}ndez}, J., {et~al.} 2016,
  \aap, 590, A32

\bibitem[{{White} {et~al.}(2007){White}, {Gabor}, \&
  {Hillenbrand}}]{White+2007}
{White}, R.~J., {Gabor}, J.~M., \& {Hillenbrand}, L.~A. 2007, \aj, 133, 2524

\bibitem[{{Yong} {et~al.}(2013){Yong}, {Mel{\'e}ndez}, {Grundahl}, {Roederer},
  {Norris}, {Milone}, {Marino}, {Coelho}, {McArthur}, {Lind}, {Collet}, \&
  {Asplund}}]{Yong+2013}
{Yong}, D., {Mel{\'e}ndez}, J., {Grundahl}, F., {et~al.} 2013, \mnras, 434,
  3542

\end{thebibliography}

\begin{appendix}

\section{Tables}
In Tables \ref{tab:GBS_abundances1} and \ref{tab:GBS_abundances2} we include the retrieved abundances and their uncertainties of the sample GBS spectra (see main text), compared with the references for the available elements.

A table with the data of all cluster stars with: star name, cluster name, number of spectra analyzed, radial velocity, Galactocentric velocity, $\teff$, $\logg$, $\vmic$, $\mathrm{[M/H]}$, bracket and differential abundances with errors is available online. We include several flags remarking any peculiarity found: (1) outlier Galactocentric velocity, (2) large radial velocity uncertainties, (3) star in the tidal tails, or (4) star used as a reference for differential abundance.

\onecolumn
\begin{landscape}

\begin{table*}
\caption{Detailed abundances of the GBS with respect to the Sun analyzed in this work (rows marked as "here") with the dispersion among the spectra quoted as uncertainty. The number of used spectra fo each star is listed in the column Nspec. The median of the individual quoted uncertainties per spectra is indicated in parenthesis. In the rows marked as  Ref" we list the reference values of the GBS abundances. The full version of the table is available at the CDS}\label{tab:GBS_abundances1}
\centering
\def\arraystretch{0.9}
\setlength\tabcolsep{2pt}
\begin{tabular}{lrrrrrrrrrrrrrrrrrr}
\hline
     Star &      & Nspec &          [CaI/H] &          [CoI/H] &          [CrI/H] &          [FeI/H] &          [MgI/H] &          [MnI/H] &          [NiI/H] &          [ScII/H] &          [SiI/H] &          [TiI/H] &           [VI/H] \\
\hline
    18Sco &  here &    24 &   $0.04\pm0.01$  &   $0.02\pm0.01$  &   $0.04\pm0.01$  &   $0.03\pm0.01$  &   $0.02\pm0.01$  &   $0.02\pm0.01$  &   $0.02\pm0.01$  &   $0.03\pm0.01$  &   $0.03\pm0.01$  &   $0.02\pm0.01$  &   $0.01\pm0.01$  \\
          &       &       &           (0.03) &           (0.03) &           (0.03) &           (0.03) &           (0.01) &           (0.05) &           (0.04) &           (0.02) &           (0.04) &           (0.04) &           (0.04) \\
          &   Ref &       &    $0.06\pm0.06$ &    $0.02\pm0.07$ &    $0.05\pm0.06$ &    $0.03\pm0.03$ &    $0.04\pm0.05$ &    $0.04\pm0.09$ &    $0.04\pm0.07$ &    $0.04\pm0.05$ &    $0.05\pm0.03$ &    $0.05\pm0.08$ &    $0.04\pm0.07$ \\
\hline
 Arcturus &  here &     5 &  $-0.55\pm0.02$  &  $-0.41\pm0.02$  &  $-0.58\pm0.01$  &  $-0.66\pm0.01$  &  $-0.26\pm0.01$  &  $-0.91\pm0.01$  &  $-0.59\pm0.01$  &  $-0.59\pm0.01$  &  $-0.37\pm0.02$  &  $-0.34\pm0.01$  &  $-0.34\pm0.02$  \\
          &       &       &           (0.07) &           (0.07) &           (0.07) &           (0.10) &           (0.04) &           (0.09) &           (0.08) &           (0.08) &           (0.06) &           (0.10) &           (0.07) \\
          &   Ref &       &   $-0.41\pm0.13$ &   $-0.41\pm0.07$ &   $-0.58\pm0.08$ &   $-0.52\pm0.08$ &   $-0.16\pm0.11$ &   $-0.89\pm0.16$ &   $-0.49\pm0.10$ &   $-0.43\pm0.15$ &   $-0.25\pm0.07$ &   $-0.31\pm0.11$ &   $-0.44\pm0.14$ \\
\hline
\end{tabular}
\end{table*}

\begin{table*}
\caption{The same as for Table~\ref{tab:GBS_abundances1} for the elements for which we do not have a reference value. The full version of the table is available at the CDS}\label{tab:GBS_abundances2}
\centering
\setlength\tabcolsep{2pt}
\begin{tabular}{lrrrrrrrrrrrrrrrrrr}
\hline
     Star & Sspec &          [AlI/H] &          [BaII/H] &          [CeII/H] &          [CuI/H] &          [EuII/H] &          [FeII/H] &          [LaII/H] &          [NaI/H] &          [NdII/H] &          [TiII/H] &           [YII/H] \\
\hline
    18Sco &    24 &   $0.04\pm0.02$  &   $0.07\pm0.01$  &   $0.04\pm0.04$  &   $0.01\pm0.01$  &   $0.16\pm0.04$  &   $0.05\pm0.01$  &  $-0.01\pm0.03$  &   $0.02\pm0.01$  &   $0.06\pm0.02$  &   $0.04\pm0.01$  &   $0.09\pm0.01$  \\
          &       &           (0.01) &           (0.04) &           (0.00) &           (0.01) &           (0.00) &           (0.04) &           (0.00) &           (0.05) &           (0.12) &           (0.04) &           (0.03) \\
\hline
 Arcturus &     5 &  $-0.23\pm0.03$  &  $-0.92\pm0.02$  &  $-0.77\pm0.04$  &  $-0.58\pm0.01$  &  $-0.39\pm0.03$  &  $-0.87\pm0.03$  &     -  &  $-0.44\pm0.02$  &  $-1.00\pm0.01$  &  $-0.47\pm0.03$  &  $-0.69\pm0.01$  \\
          &       &           (0.01) &           (0.03) &           (0.00) &           (0.00) &           (0.00) &           (0.04) &            - &           (0.02) &           (0.23) &           (0.08) &           (0.04) \\
\hline
\end{tabular}
\end{table*}

\begin{table*}
\small
\caption{Details of the cluster stars. We list the cluster, Gaia DR2 source id, number of spectra analyzed, radial velocity, Galactic velocities, atmospheric parameters ($\teff$, $\logg$, $\vmic$, $\mathrm{[M/H]}$), and bracket and differential abundances with errors. We include several flags remarking any peculiarity found. The full version of the table is available at the CDS.}\label{tab:all_star_abus}
\centering
\setlength\tabcolsep{3pt}
\begin{tabular}{ccccccccccccccc}
\hline
Cluster & Star                & Num & \vr         & ($U,V,W$)         & $\teff$     & $\logg$       & $\vmic$       & $\mh$         & [Fe/H]        & $\delta$Fe   & ... & [Eu/H] & $\delta$Eu   & Flags$^{\star}$ \\
        &                     & Spec& (\kms)      & (\kms)            & (K)         & (dex)         & (\kms)        & (dex)         & (dex)         & (dex)        & &  (dex)        & (dex)       &   \\
\hline
Hyades  & 3312575685471393664 & 3   & $40.3\pm0.1$&($-42.1,19.5,-1.6$)& $5897\pm22$ & $4.45\pm0.04$ & $1.33\pm0.06$ & $0.00\pm0.02$ & $0.11\pm0.02$ & $0.0\pm0.01$ & ... & $0.06\pm0.50$ &$ -0.01\pm0.50$ & R\\
\hline
\end{tabular}
\flushleft Flags: O = outlier in ($U,V,W$); E = large uncertainty in \vr; T = star in the tidal tails; R = star used as reference for the computation of differential abundances
\end{table*}

\end{landscape}

\section{Additional figures}

\begin{figure*}
\centering
\includegraphics[width=0.8\textwidth]{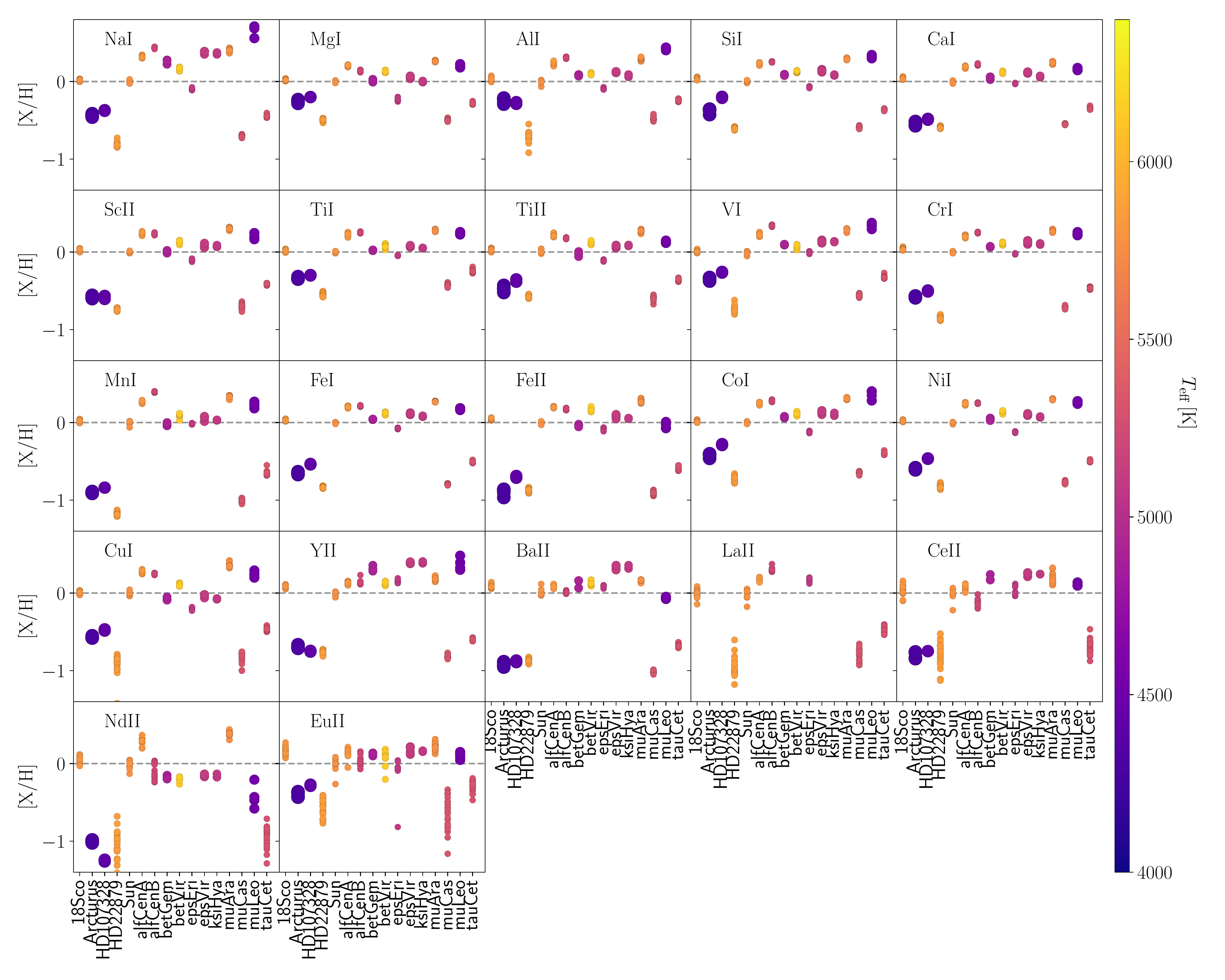}
\caption{Abundances with respect to the mean solar abundance obtained in this work for the several spectra of the analyzed GBS (the stars are labeled by their names along the $x$-axis in alphabetical order). Color code corresponds to effective temperature, and symbol size is scaled by the surface gravity, where larger sizes correspond to giant stars (smaller gravity).}\label{fig:abu_GBS_spectr}
\end{figure*}

\begin{figure*}
\centering
\includegraphics[width=0.8\textwidth]{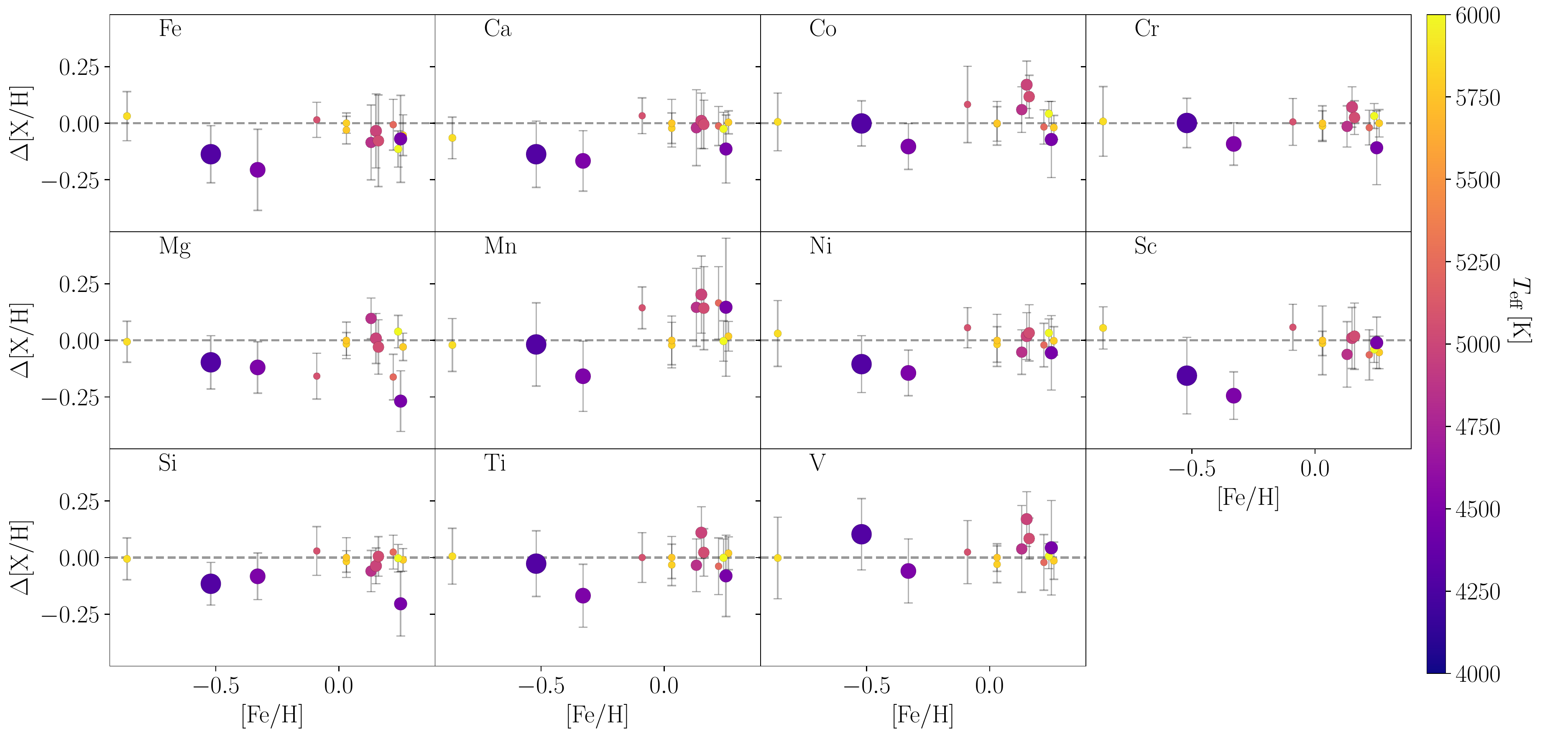}
\caption{Differences (here -- reference) in abundances [X/H] for the selection of the \gls{GBS}. We only plot the elements for which a reference value exists in \citet{Jofre+2015}. Error bars correspond to the quadratic sum of the quoted errors in this work and in the reference. }\label{fig:abu_GBS}
\end{figure*}

\begin{figure*}
\centering
\includegraphics[width=\textwidth]{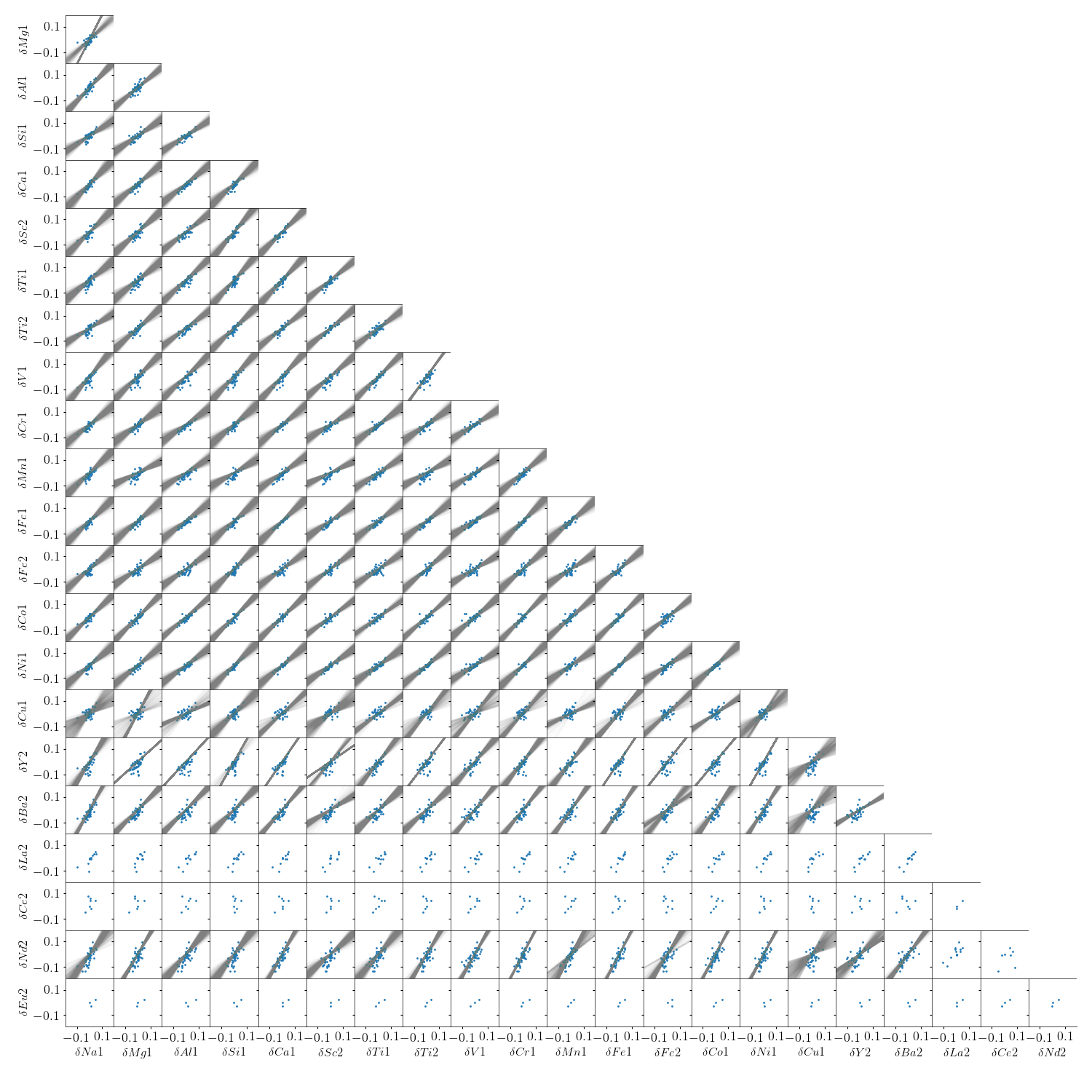}
\caption{Differential abundances $\delta X$, $\delta Y$ for all possible combination of elements in the Hyades stars.  Grey lines represent the results and uncertainties of linear fits performed as described in the main text, sampling the posterior distribution.}\label{fig:XHXH_Hyades}
\end{figure*}

\begin{figure*}
\centering
\includegraphics[width=\textwidth]{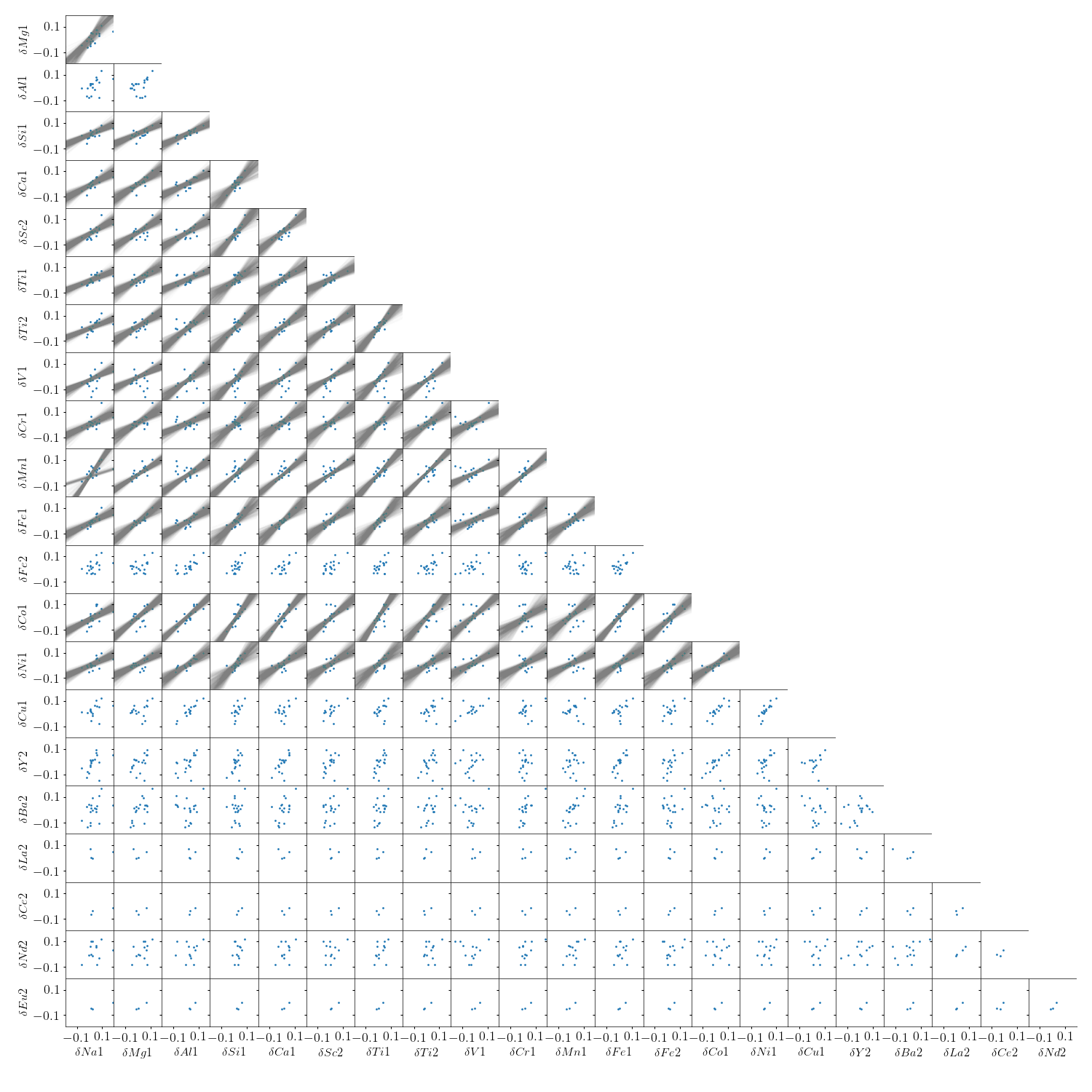}
\caption{The same as Fig.~\ref{fig:XHXH_Hyades}, for NGC~2632.}\label{fig:XHXH_NGC2632}
\end{figure*}

\begin{figure*}
\centering
\includegraphics[width=\textwidth]{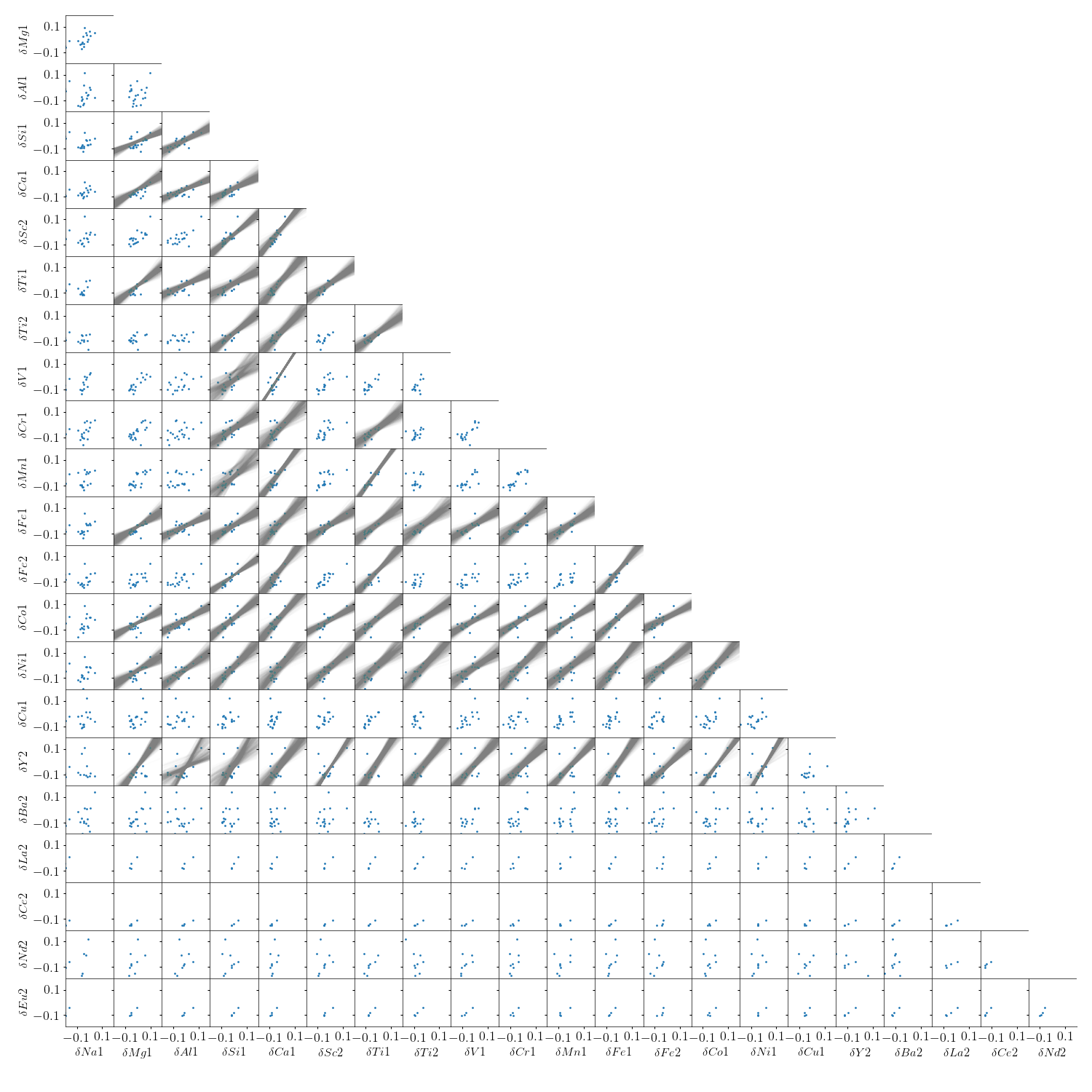}
\caption{The same as Fig.~\ref{fig:XHXH_Hyades}, for Ruprecht~147.}\label{fig:XHXH_Rup147}
\end{figure*}

\end{appendix}

\end{document}